\documentclass[12pt]{article}
\usepackage{geometry}                
\usepackage{graphicx}
\usepackage{amssymb}
\usepackage{float}

\usepackage[braket]{qcircuit}

\usepackage[parfill]{parskip} 

\usepackage{booktabs} 
\usepackage{array} 
\usepackage{paralist} 
\usepackage{verbatim} 
\usepackage{subfig} 

\usepackage{tikz}
\usetikzlibrary{shapes}
\usetikzlibrary{circuits}
\usetikzlibrary{circuits.logic.US}

\usetikzlibrary{arrows}

\usetikzlibrary{positioning}

\usepackage{verbatim}

\usepackage[matrix,frame,arrow]{xypic}

\usepackage{datetime}

\usepackage{fancyhdr} 
\usepackage{sectsty}

\usepackage[nottoc,notlof,notlot]{tocbibind} 
\usepackage[titles,subfigure]{tocloft} 

\usepackage{blochsphere}

\def \bmatrix#1{ \left [ \matrix{#1}  \right ] }

\title{{\bf Loading Classical Data into\\a Quantum Computer}\\{\normalsize\vspace*{0.6in}by}}
\author{John A. Cortese\footnote{Corresponding author: jo15811@mit.edu\vspace*{0.2in}} $\;$ and $\,$Timothy M. Braje\vspace*{0.6in}\\Lincoln Laboratory\\Massachusetts Institute of Technology\\Lexington, Massachusetts, USA}

\emergencystretch=\maxdimen
\renewcommand{\arraystretch}{1.25}

\date{}
\begin{document}
\maketitle

\begin{tikzpicture}[remember picture,overlay]
\node [yshift=1.4in] at (current page.south) [text width=6.5in] {DISTRIBUTION STATEMENT A. Approved for public release: distribution unlimited.\\\vspace*{0.08in}This material is based upon work supported under Air Force Contract No. FA8721-05-C-0002 and/or FA8702-15-D-0001. Any opinions, findings, conclusions or recommendations expressed in this material are those of the author(s) and do not necessarily reflect the views of the U.S. Air Force. \textcopyright 2017 Massachusetts Institute of Technology.};
\end{tikzpicture}

\thispagestyle{fancy} 

\ifdefined\jacflag
\settimeformat{ampmtime}
\begin{center}
Time: \currenttime
\end{center}
\fi

\newpage
\lhead{}\chead{}\rhead{Page \thepage}
\tableofcontents

\newpage
\listoffigures

\newpage
\listoftables

\newpage

\section{Overview}
\label{invention}

This document describes a family quantum circuits and associated techniques and design principles which collectively serve to efficiently transfer data, specifically binary data or bits, from the classical domain (classical world) into the quantum domain (quantum world). In addition, the quantum data is formatted into quantum states for follow-on processing by efficient quantum algorithms.

\subsection{Block diagram description of a quantum algorithm}

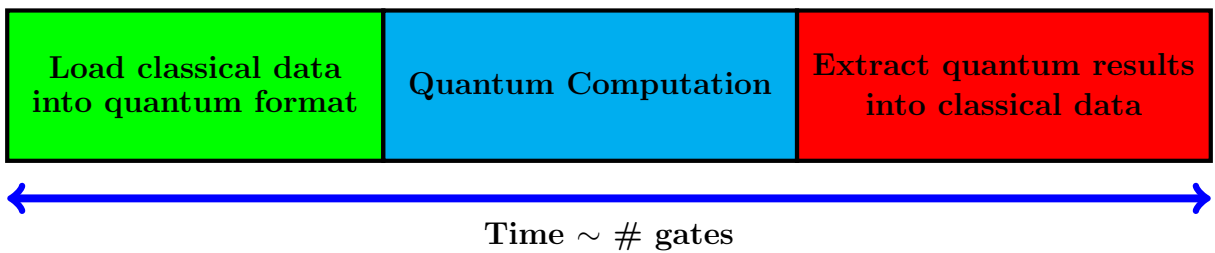
\begin{figure}[h!]
\begin{center}
\begin{tikzpicture}
\draw [fill=green,ultra thick] (-12,-1) rectangle (-7,1);
\node at (-9.5,0.25) {\bf Load classical data};
\node at (-9.5,-0.25) {\bf into quantum format};
\draw [fill=cyan,ultra thick] (-7,-1) rectangle (-1.5,1);
\node at (-4.25,0) {\bf Quantum Computation};
\draw [fill=red,ultra thick] (-1.5,-1) rectangle (4,1);
\node at (1.25,0.25) {\bf Extract quantum results};
\node at (1.25,-0.25) {\bf into classical data};
\node at (-4,-2) {\bf Time $\sim$ \# gates};
\draw [<->,blue,ultra thick,line width = 3pt] (-12,-1.5) -- (4,-1.5);
\end{tikzpicture}
\caption[Stages of execution in a quantum computer]{The three stages of execution in a quantum computer.}
\label{fig3}
\end{center}
\end{figure}

The three generic stages of execution in a quantum computer are shown in Figure \ref{fig3}. The work in this document concerns the leftmost, green stage 
shown in Figure \ref{fig3} labeled {\em Load classical data into quantum format}.

\section{The data loading quantum circuits}

Quantum circuits for loading classical data into quantum states for processing by a quantum computer are the focus of this document. As different quantum algorithms have varying constraints on how the classical input data 
is loaded and formatted into the corresponding quantum states, a number of data loading circuits will be presented. 
The utility and
usage of each data loading circuit is discussed in the context of the quantum algorithms to which the classical 
data is being passed.

\subsection{Quantum state formats for representing classical data}

Consider the problem of mapping a matrix of classical single bit values ($b_{ij}$) into a quantum state. 
As an example, take the $2$ by $2$ matrix $A=
\bmatrix{b_{00} \quad b_{01} \cr b_{10} \quad b_{11} }$. 
Note that for bit $b_{ij}$, the {\it i} subscript indicates the row, with {\it i = 0 or 1}, and the {\it j} subscript indicates the column, with
{\it j = 0 or 1}. 
The corresponding quantum state which will hold the matrix $A$ bit values is $\psi_A$. The quantum state $\psi_A$ 
uses three (3) qubits to represent the four single bit values $b_{ij}$, as well as the corresponding positions of the bits in the matrix. 
\begin{equation}
|\psi_A\rangle\; =\; |00\rangle\otimes |b_{00}\rangle \;+\;  |01\rangle\otimes |b_{01}\rangle \;+\;  |10\rangle\otimes |b_{10}\rangle \;+ \; |11\rangle\otimes |b_{11}\rangle 
\label{Eqn1}
\end{equation}
$$
\equiv 
\; |00 b_{00}\rangle \;+\;  |01b_{01}\rangle \;+\;  |10b_{10}\rangle \;+ \; |11b_{11}\rangle .
$$
The symbol $\otimes$ is the tensor product operator and will be used to delineate groups of qubits within a quantum state. Quantum information notation often drops the tensor symbol to write the state as shown in the second line of Equation 
\ref{Eqn1}. The first, leftmost qubit of the state represents the row within the matrix, corresponding to the {\it i} index. Similarly the second qubit represents the column, corresponding to the index {\it j}. The third, rightmost qubit represents the single bit value of the corresponding (row,column) matrix entry. 
As is traditional in quantum information, the overall quantum state normalization constant is dropped for readability. The overall state normalization constant is straightforward to compute and reinsert when necessary. For the state in Equation \ref{Eqn1}, the normalization constant is $\frac{1}{2}$, regardless of the values of the $b_{ij}$.  

This document describes several families of data loading circuits.
Each family is optimal under a different set of constraints. Each circuit family is given in order of circuit complexity, with a
corresponding discussion. The material in this document is concerned with the gate based model of quantum computation
and the circuits are presented in a gate based framework.

\subsection{Circuit Family \#1}

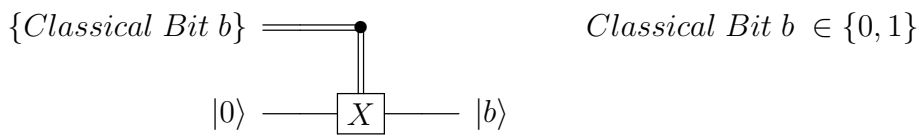
\begin{figure}[h!]
\begin{center}
\[
\Qcircuit @C=1.2em @R=2.0em {
 \lstick{\{Classical \;Bit \;b\} } & \cw & \control \cwx[1] \cw & & & & & \rstick{Classical \;Bit \;b\; \in \{0,1\}} \\
\lstick{|0\rangle} & \qw & \gate{X}  & \qw & \rstick{|b\rangle} \qw & & & & & & & & \\
}
\]
\caption{Loading a classical bit {\em b} into the qubit quantum state $|b\rangle$.}
\label{figb}
\end{center}
\end{figure}

The quantum circuit shown in Figure \ref{figb} loads a single classical bit into a qubit. 
In Figure \ref{figb},  
double wires are classical wires conveying a classical bit value = \{0,1\}. 
Single wires are quantum wires along which quantum states or qubits move. 
The box with the {\bf X} inside is a quantum bit flip gate. The quantum bit flip gate
acts the same on quantum states as a classical inverter gate does on classical bits, reversing the value of the qubit. 
In both the classical and quantum scenario bit flip gate action, a "0" goes to a "1" and a "1" goes to a "0". 
In the circuit shown in Figure \ref{figb},
the quantum bit flip gate is a controlled gate. The control is a classical wire feeding into the top of the gate. 
The solid dot indicates the control wire 
for the corresponding box/gate action. If the classical control wire is a "0", then the bit flip gate is not executed.
If the classical control wire is a "1", then the bit flip gate is executed.
The circuit diagram data 
flow is always left to right in quantum circuits. 

\begin{figure}[h!]
\begin{center}
\[
\Qcircuit @C=1.2em @R=2.0em {
 \lstick{\{Bit = 0\} } & \cw & \control \cwx[1] \cw & & & & & \rstick{\underline{Bit \;flip \;gate \;is \;not \;executed.}} \\
\lstick{|0\rangle} & \qw & \gate{X}  & \qw & \rstick{|0\rangle} \qw & & & & & & & & \\
}
\]
\caption{Loading a classical "0" bit into the $|0\rangle$ qubit quantum state.}
\label{figc}
\end{center}
\end{figure}
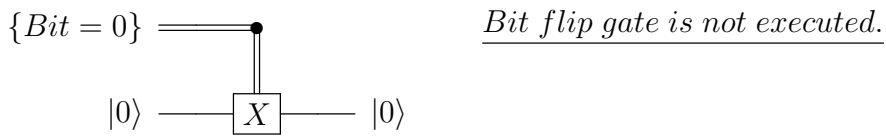

\vspace*{0.3in}
\begin{figure}[h!]
\begin{center}
\[
\Qcircuit @C=1.2em @R=2.0em {
 \lstick{\{Bit = 1\} } & \cw & \control \cwx[1] \cw & & & & & \rstick{\underline{Bit \;flip \;gate \;is \;executed.}} \\
\lstick{|0\rangle} & \qw & \gate{X}  & \qw & \rstick{|1\rangle} \qw & & & & & & & & \\
}
\]
\caption{Loading a classical "1" bit into the $|1\rangle$ qubit quantum state.}
\label{figd}
\end{center}
\end{figure}
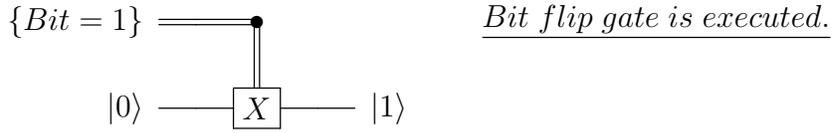

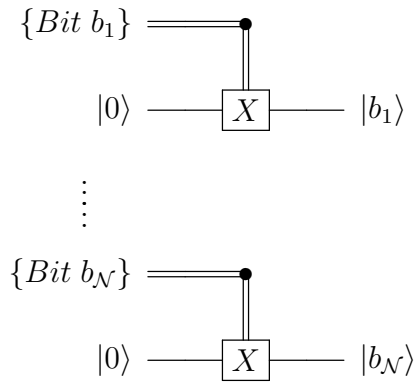
\begin{figure}[h!]
\begin{center}
\[
\Qcircuit @C=1.2em @R=2.0em {
 \lstick{\{Bit \;b_1\} } & \cw & \control \cwx[1] \cw \\
\lstick{|0\rangle} & \qw & \gate{X}  & \qw & \rstick{|b_1\rangle} \qw & & & & & & & & \\
}
\]
$$
\vdots \qquad \qquad \qquad  \qquad \qquad \qquad  \qquad \qquad \qquad  \qquad
$$
\vspace*{-0.4in}
$$
\vdots \qquad \qquad \qquad  \qquad \qquad \qquad  \qquad \qquad \qquad  \qquad
$$
\[
\Qcircuit @C=1.2em @R=2.0em {
 \lstick{\{Bit \;b_\mathcal{N}\} } & \cw & \control \cwx[1] \cw \\
\lstick{|0\rangle} & \qw & \gate{X}  & \qw & \rstick{|b_\mathcal{N}\rangle} \qw & & & & & & & & \\
}
\]
\caption[Loading $\mathcal{N}$ bits into $\mathcal{N}$ qubits in a single gate depth quantum circuit]{Loading $\mathcal{N}$ classical bits $\{ b_1,\cdots,b_\mathcal{N}\}$ into $\mathcal{N}$ qubits in a quantum circuit with a gate depth equal to one.}
\label{fige}
\end{center}
\end{figure}

In circuit family \#1, each classical bit is stored in one qubit and requires the execution of one quantum gate to implement the classical bit loaded into a qubit storage operation. 
To better understand the asymptotic behavior of the size of the circuits and other aspects of the circuit families,
let the total number of classical 
bits being loaded into
the quantum computer be $\mathcal{N}$. 
If the input data consists of $N$ words where each word is $P$ bits long, 
then the total number of bits $\mathcal{N}$ are $\{\,b_i\,\}\;\in \{0,1\}$ with
$i\;=\;1,\,2,\,\cdots,\,\mathcal{N}\,=\,N\,P$.
The circuit shown in Figure \ref{fige} which is loading the $\mathcal{N}=NP$ classical bits will  
require a quantum state consisting of $\mathcal{N}=NP$ qubits to store these bits. Computationally, the circuit requires 
the execution of $\mathcal{N}=NP$ gates in parallel. The gate depth corresponds to the time the circuit will take to execute on the input data. The gate depth of the quantum circuit in Figure \ref{fige}  
is $1$. The gate depth for all family \#1 data loading quantum circuits is $1$. 

The $\mathcal{N}$ qubit 
quantum state $\psi$ produced by the circuit is 

\begin{equation}
\psi \;=\;  | b_1\rangle \otimes | b_2\rangle \otimes \cdots| b_{\mathcal{N}}\rangle\;\equiv\; | b_1\, b_2\, \cdots\, b_{\mathcal{N}}\,\rangle.
\label{Ea}
\end{equation}

A table of the resource requirements for each data loading circuit
family 
will gradually be compiled. For circuit family \#1 described in this section, the resource requirements are given in Table \ref{table1}.

\begin{table}[h!]
\begin{center}
\begin{tabular}{||c | c | c |c| c|c||}
\hline \hline 
 &Number & Number of & Number& & Number\\
 Circuit & of & qubits in& of &  Gate& of \\
 family&classical&the quantum&total & depth &ancilla \\
  &  bits  &state $\psi$&gates & &qubits \\
\hline 
\#1 & $\mathcal{N}$ & $\mathcal{N}$ & $\mathcal{N}$ & $1$ &$0$\\
\hline \hline  
\end{tabular}
\end{center}
\caption[Resource requirements for data loading Circuit Family \#1]{Resource requirements for data loading circuit family \#1. $N$ is the number of vector entries or words. Let $N=2^n$. $P$ is the number of bits per vector entry or word. The total number of classical bits is $\mathcal{N}\;=\;N\, P\;=\; 2^n\,P$.}
\label{table1}
\end{table} 

The quantum state shown in Equation \ref{Ea} is not optimal for use as the input to a quantum algorithm exhibiting exponential speedup. Loading $\mathcal{N}$ classical bits into a quantum state composed
of $Log_2(\,\mathcal{N}\,)$ or fewer qubits is needed. 
This fact motivated the development of circuit family \#2. 

\subsection{Circuit Family \#2}

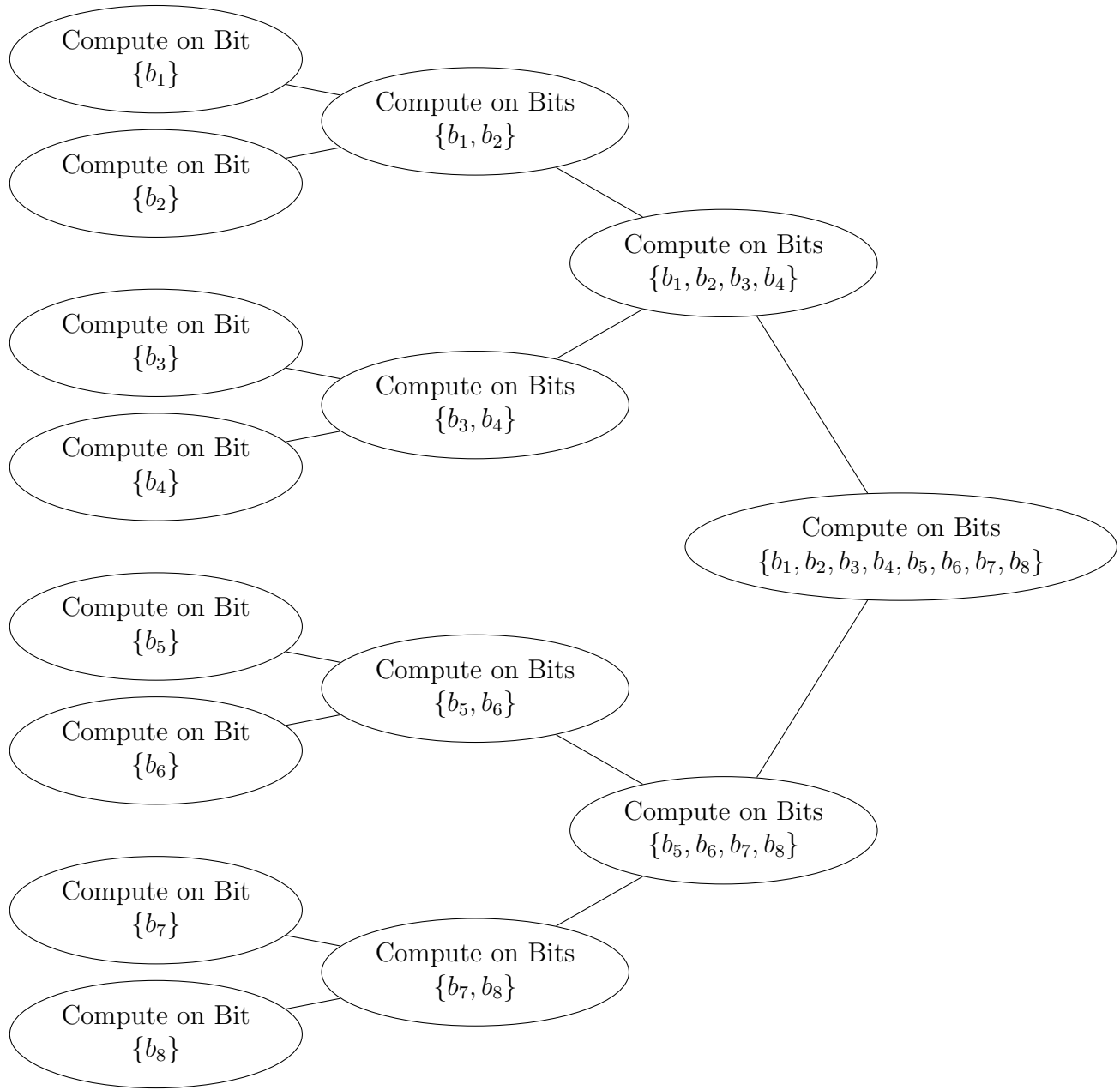
\begin{figure}[h!]
\hspace*{-0.1in}
\begin{tikzpicture}[scale=0.55,level 1/.style={sibling distance=160mm,level distance=50mm},level 2/.style={sibling distance=80mm,level distance=70mm},level 3/.style={sibling distance=35mm,level distance=90mm}]
 \node[align=center,ellipse,draw] {Compute on Bits\\\{$b_1,b_2,b_3,b_4,b_5,b_6,b_7,b_8$\}} [grow=left] 
	child {node[align=center,ellipse,draw] {Compute on Bits\\\{$b_1,b_2,b_3,b_4$\}}
	child {node[align=center,ellipse,draw] {Compute on Bits\\\{$b_1,b_2$\}}
	child {node[align=center,ellipse,draw] {Compute on Bit\\\{$b_1$\}}}
	child {node[align=center,ellipse,draw] {Compute on Bit\\\{$b_2$\}}}
	}
	child {node[align=center,ellipse,draw] {Compute on Bits\\\{$b_3,b_4$\}}
	child {node[align=center,ellipse,draw] {Compute on Bit\\\{$b_3$\}}}
	child {node[align=center,ellipse,draw] {Compute on Bit\\\{$b_4$\}}}
	}
	}
	child {node[align=center,ellipse,draw] {Compute on Bits\\\{$b_5,b_6,b_7,b_8$\}}
	child {node[align=center,ellipse,draw] {Compute on Bits\\\{$b_5,b_6$\}}
	child {node[align=center,ellipse,draw] {Compute on Bit\\\{$b_5$\}}}
	child {node[align=center,ellipse,draw] {Compute on Bit\\\{$b_6$\}}}
	}
	child {node[align=center,ellipse,draw] {Compute on Bits\\\{$b_7,b_8$\}}
	child {node[align=center,ellipse,draw] {Compute on Bit\\\{$b_7$\}}}
	child {node[align=center,ellipse,draw] {Compute on Bit\\\{$b_8$\}}}
	}
	}; 
\end{tikzpicture}
\caption[Recursive Computation on $\mathcal{N}=8$ bits]{Recursive Computation on $\mathcal{N}=8$ bits.}
\label{figtree}
\end{figure}

Circuit family \#1 loads $\mathcal{N}$ bits into a quantum state of size $\mathcal{N}$ using a gate depth of $1$.  
The benefit of quantum circuits over classical circuits is the potential for the quantum circuit to exhibit 
exponential speedup in execution time, which translates to a logarithmic reduction in the gate depth. In 
order for this speedup to be possible, the size, meaning the number of qubits of the quantum state 
containing the relevant classical data, should be logarithmic 
in the number of classical bits being manipulated.
That is, if there is no pattern or symmetry indicating preference among the incoming data items, then all the data must be considered equally. A circuit implementing a generic algorithm on $\mathcal{N}$ data items should
have a gate depth which is logarithmic in $\mathcal{N}$. For the case of $\mathcal{N}\,=\,8$ bits, the recursive exploration of the data is shown as a tree in Figure \ref{figtree}.

The size of the circuit family \#1 quantum state is $\mathcal{N}$ qubits, which is typically too large for quantum circuitry 
to use and still exhibit exponential speedup over classical circuitry. It is possible to pack $\mathcal{N}$ classical bits into a 
quantum state of size $Log_2(\mathcal{N})$ qubits. Such a compression of the classical bit count is a key factor 
enabling quantum algorithm design for exponential speedup over classical algorithms. Circuit family \#2 exhibits this logarithmic compression of classical 
bits into qubits. As a result, circuit family \#2 is an enabling technology for the implementation of quantum 
algorithms exhibiting exponential speedup over classical algorithms.

\begin{figure}[h!]
\begin{center}
\[
\Qcircuit @C=1.0em @R=2em {
\lstick{|0\rangle} & \gate{H} & \qw & \qw &     \ctrl{1}   & \qw & \qw  & \qw & \qswap  \qwx[3]  & \qw & \qw & \qw &\rstick{\beta}\\
 \lstick{\ket{b_{00}}} & \qw  & \qw & \qw   & \qswap \qwx[1] & \qw  & \qw & \qw  & \qw & \qswap  \qwx[3] &  \qw  & \qw &\rstick{\gamma}\\
 \lstick{\ket{b_{01}}} & \qw & \qw & \qw  & \qswap \qw  &\rstick{\hspace*{10em}Discard \;this\;qubit} \\
 \lstick{\ket{0}}  & \gate{H}  & \qw & \qw  &  \ctrl{1} & \qw & \qw & \qw    & \qswap   &\rstick{\hspace*{5.5em}Discard \;this\;qubit} \\
 \lstick{\ket{b_{10}}} & \qw  & \qw     & \qw & \qswap \qwx[1] & \qw & \qw & \qw  & \qw & \qswap   &\rstick{\hspace*{4.5em}Discard \;this\;qubit}\\
 \lstick{\ket{b_{11}}} & \qw    & \qw & \qw & \qswap \qw   &\rstick{\hspace*{10em}Discard \;this\;qubit}\\
 \lstick{\ket{0}}  & \gate{H} & \qw & \qw & \qw & \qw &  \qw & \qw &  \ctrl{-3} \qw  &  \ctrl{-2} \qw &  \qw & \qw &\rstick{\alpha} \\
}
\]
\caption[A Circuit Family \#2 layout]{A circuit family \#2 implementation layout loading the four classical bits, $\{\, b_{00}, \,b_{01}, \,b_{10}, \,b_{11}\,\}$ into the three qubits $|\alpha \beta \gamma\rangle$. Recall that a classical bit $b$ is loaded into a single qubit denoted $|b\rangle$ using the quantum circuit/gate shown in Figure \ref{figb}.}
\label{figa}
\end{center}
\end{figure}
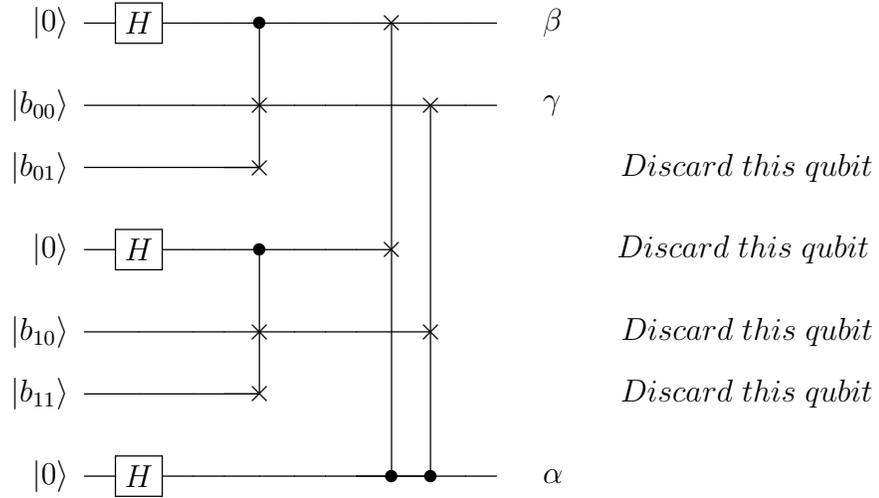

The three qubit state $|\psi_A\rangle$ on the right hand side of the circuit in Figure \ref{figa} is storing the four classical bit values of the matrix $A$.

\begin{equation}
|\psi_A\rangle \;=\; 
|\psi_{\alpha\beta\gamma}\rangle \;=\; |\alpha\beta\gamma\rangle
\;
=\; |00\rangle \otimes |b_{00} \rangle\;+\;  |01\rangle \otimes |b_{01}\rangle \;+\;  |10\rangle \otimes |b_{10}\rangle\;+\;  |11\rangle \otimes| b_{11}\rangle.
\label{Eb}
\end{equation}

Key aspects of circuit family \#2 are shown in Figure \ref{figa} and will be described in the next several sections. 

\subsection{More about quantum circuits}

To understand the circuit shown in Figure \ref{figa},  more detail about the constituent gates must be given. 
Quantum circuits are composed of quantum gates. Quantum gates act linearly on their input. 
Therefore, knowing the action of any gate on all possible computational basis inputs suffices 
to completely characterize the gate. 
In this section, several gates will be described which are used in the circuit shown in Figure \ref{figa}. 

\subsubsection{The quantum Swap gate}

\begin{figure}[H]
\begin{center}
\[
\Qcircuit @C=1.0em @R=2em {
 \lstick{\ket{\alpha}} & \qw  & \qw & \qswap \qwx[1] & \qw  & \qw & \rstick{\ket{\beta}} \\
 \lstick{\ket{\beta}} & \qw  & \qw & \qswap \qw & \qw  & \qw   & \rstick{\ket{\alpha}} \\
}
\]
\caption[The quantum Swap gate]{The quantum Swap gate interchanges the states of two qubits.}
\label{fig12}
\end{center}
\end{figure}
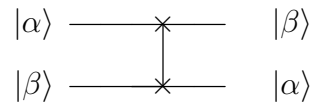

The quantum swap gate shown in Figure \ref{fig12}  will be used extensively in the circuits to follow. 

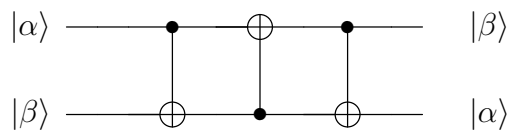
\begin{figure}[h!]
\begin{center}
\[
\Qcircuit @C=1.0em @R=2em {
 \lstick{\ket{\alpha}} & \qw  & \qw & \ctrl{1} & \qw & \targ \qwx[1] \qw & \qw& \ctrl{1} &\qw  &\qw & \rstick{\ket{\beta}} \\
 \lstick{\ket{\beta}} & \qw  & \qw & \targ \qw & \qw & \control \qw & \qw &\targ \qw &\qw  &\qw   & \rstick{\ket{\alpha}} \\
}
\]
\caption[A Swap gate implementation using CNOT's]{The Swap gate implemented with three Controlled-Not (CNOT) gates.}
\label{figg}
\end{center}
\end{figure}

The quantum swap gate shown in Figure \ref{fig12} can be implemented with three Controlled-Not (CNOT) gates as shown in Figure \ref{figg}. 

\subsubsection{The Controlled Swap gate}

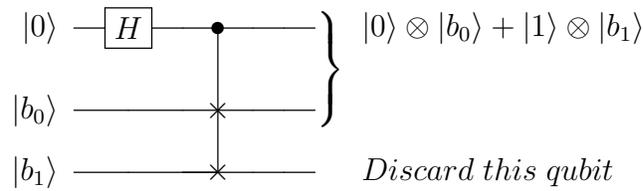
\begin{figure}
\begin{center}
\[
\Qcircuit @C=1.0em @R=2em {
\lstick{|0\rangle} & \gate{H} &\qw &  \ctrl{1} & \qw & \qw & \qw& \rstick{ |0\rangle \otimes |b_0\rangle  + |1\rangle \otimes |b_1\rangle } \\
 \lstick{\ket{b_0}} & \qw  & \qw & \qswap \qwx[1] & \qw & \qw  & \qw   \gategroup{1}{6}{2}{7}{1em}{\}}\\
  \lstick{\ket{b_1}} & \qw  & \qw & \qswap \qw & \qw  & \qw &\qw & \rstick{Discard\;this\;qubit}\\
}
\]
\caption{The Controlled Swap (CSwap) gate.}
\label{fig14}
\end{center}
\end{figure}

The controlled quantum swap gate (C-Swap or CS) is shown in Figure \ref{fig14}. Whether two qubit states are swapped depends on the quantum state of a control line. 
The three qubit input state to the quantum circuit in Figure \ref{fig14} is $|\, q_1 \, q_2 \, q_3 \,\rangle \;=\; |\,  0 \, b_0 \, b_1 \,\rangle$. The three qubit output state at the right hand side of the circuit in Figure \ref{fig14} is the superposition shown in Equation \ref{eswap1}.

\begin{equation}
|\, q_1 \, q_2 \, q_3 \,\rangle \;=\; | \, 0 \, b_0 \, b_1 \,\rangle \; +\; | \, 1 \, b_1 \, b_0 \,\rangle,
\label{eswap1}
\end{equation}

where, as discussed previously, without loss of generality, the normalization coefficient, 
which in this case is $\frac{1}{\;\sqrt{2}\;}$, has been dropped from Equation \ref{eswap1}.

\subsubsection{The Toffoli gate}

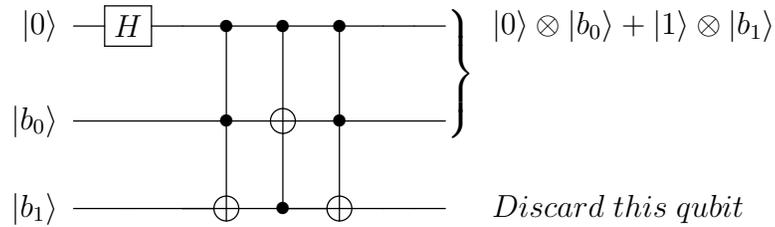
\begin{figure}[H]
\begin{center}
\[
\Qcircuit @C=1.0em @R=2em {
\lstick{|0\rangle} & \gate{H} &\qw &  \ctrl{1} &  \ctrl{1}&  \ctrl{1}& \qw & \qw & \qw& \rstick{ |0\rangle \otimes |b_0\rangle  + |1\rangle \otimes |b_1\rangle } \\
 \lstick{\ket{b_0}} & \qw  & \qw & \ctrl{1} & \targ \qwx[1] & \ctrl{1}& \qw & \qw  & \qw   \gategroup{1}{8}{2}{9}{1em}{\}}\\
 \lstick{\ket{b_1}} & \qw  & \qw & \targ \qw & \control \qw & \targ \qw& \qw  & \qw &\qw & \rstick{Discard \;this\;qubit}\\
}
\]
\caption[The CSwap gate implemented using Controlled Controlled Nots]{The controlled quantum swap gate using Controlled Controlled Nots (CCNot's), which are equivalent to Toffoli gates.}
\label{figh}
\end{center}
\end{figure}

The Toffoli gate is a Controlled-Controlled-Not gate (CCNot). It has two control lines and executes a bit flip on a third qubit if and only if the two control qubits are both in the $|1\rangle$ state. A controlled swap gate can be implemented with three Toffoli gates as shown in Figure \ref{figh}. 


\begin{figure}[h!]
\begin{center}\[
\Qcircuit @C=1.0em @R=2em {
\lstick{|0\rangle} & \gate{H} &\qw &  \qw &  \ctrl{1}&  \qw & \qw & \qw & \qw& \rstick{ |0\rangle \otimes |b_0\rangle  + |1\rangle \otimes |b_1\rangle } \\
 \lstick{\ket{b_0}} & \qw  & \qw & \ctrl{1} & \targ \qwx[1] & \ctrl{1}& \qw & \qw  & \qw   \gategroup{1}{8}{2}{9}{1em}{\}}\\
 \lstick{\ket{b_1}} & \qw  & \qw & \targ \qw & \control \qw & \targ \qw& \qw  & \qw &\qw & \rstick{Discard \;this\;qubit}\\
}
\]
\caption[The CSwap gate implemented using CCNot and CNOT gates]{The controlled quantum swap gate using one CCNot (Toffoli) gate and two Controlled-Not gates.}
\label{figi}
\end{center}
\end{figure}
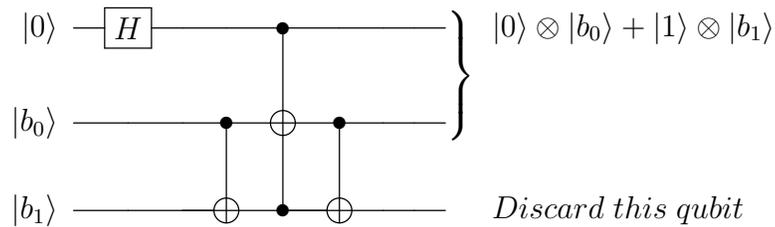

A simplification of the circuit in Figure \ref{figh} using only one Toffoli gate and two Controlled-Not gates is shown in Figure \ref{figi}.
A Toffoli gate can be decomposed into a sequence of one and two qubit gates. The circuit shown in Figure \ref{figk} is one such decomposition. 
The single qubit  gate $S$ in Figure \ref{figk} is the Phase gate, defined as
$S=\bmatrix{1 \quad 0 \cr 0 \quad i}$,
where $i$ is the square root of $-1$, namely $i\;=\;\sqrt{\,-1\,}$. 

\begin{center}
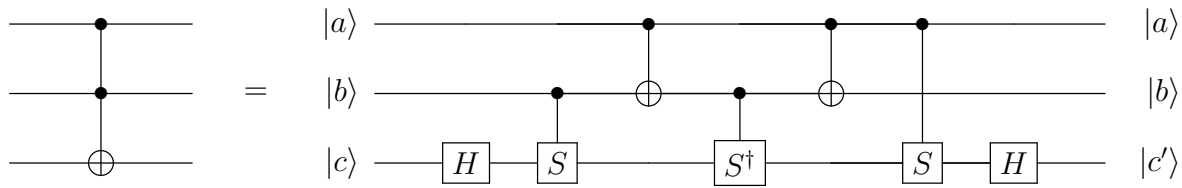
\begin{figure}
\[
\Qcircuit @C=1.3em @R=.6em @! {
 &  \ctrl{1} & \qw  &    &    \lstick{\ket{a}}  &\qw & \qw & \ctrl{1} \qw &
         \qw & \ctrl{1} \qw & \ctrl{2} \qw  & \qw &
         \qw & \lstick{\ket{a}}\\
  & \ctrl{1} &  \qw & \lstick{=}  &    \lstick{\ket{b}} &\qw   & \ctrl{1} & \targ \qw &
         \ctrl{1} \qw  & \targ \qw & \qw & \qw & \qw &  \lstick{\ket{b}}\\
 & \targ & \qw     &  & \lstick{\ket{c}} & \gate{H} & \gate{S} & \qw &  \gate{S^\dagger} & \qw &
         \gate{S} \qw & \gate{H} \qw & \qw & \lstick{\ket{c^\prime}}
}
\]
\caption[An implementation of the CCNot (Toffoli) gate]{A two qubit gate implementation of the three qubit Toffoli (CCNot) gate.}
\label{figk}
\end{figure}
\end{center}

\subsubsection{Controlled Swaps of multiple qubits}

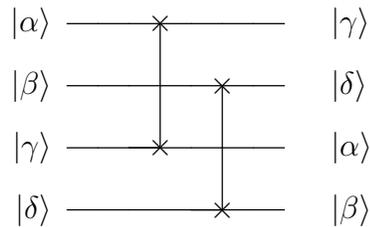
\begin{figure}[h!]
\begin{center}
\[
\hspace*{-0.5in}
\Qcircuit @C=1.0em @R=2em {
 \lstick{\ket{\alpha}} & \qw  & \qw & \qswap \qwx[2] &\qw  & \qw  & \qw  & \qw & \rstick{\ket{\gamma}}\\
 \lstick{\ket{\beta}} & \qw  & \qw &  \qw & \qw & \qswap \qwx[2]   & \qw & \qw & \rstick{\ket{\delta}} \\
  \lstick{\ket{\gamma}} & \qw  & \qw & \qswap \qw & \qw& \qw & \qw  & \qw & \rstick{\ket{\alpha}} \\
 \lstick{\ket{\delta}} & \qw  & \qw & \qw & \qw & \qswap \qw & \qw  & \qw & \rstick{\ket{\beta}} \\
}
\]
\caption[Quantum circuit swapping two pairs of qubits]{Using two Swap gate's to exchange two pairs of two qubits.}
\label{figf}
\end{center}
\end{figure}

Looking back at the recursive structure shown in Figure \ref{figa}, one notes the need for a quantum circuit which will swap into
superposition more than two bits. 
Building upon the one qubit circuit shown in Figure \ref{fig14}, a two qubit swap quantum circuit is shown in Figure \ref{figf}.

\begin{figure}[h!]
\begin{center}
\[
\hspace*{-0.5in}
\Qcircuit @C=1.0em @R=2em {
\lstick{|0\rangle} & \gate{H} &\qw &  \ctrl{1} & \qw &  \ctrl{2} & \qw & \qw& \rstick{ |0\rangle \otimes |b_0b_1\rangle  + |1\rangle \otimes |b_2b_3\rangle } \\
 \lstick{\ket{b_0}} & \qw  & \qw & \qswap \qwx[2] &\qw  & \qw  & \qw  & \qw \gategroup{1}{7}{3}{8}{1em}{\}}\\
 \lstick{\ket{b_1}} & \qw  & \qw &  \qw & \qw & \qswap \qwx[2]   & \qw & \qw \\
  \lstick{\ket{b_2}} & \qw  & \qw & \qswap \qw & \qw& \qw & \qw  & \qw & \rstick{Discard \;these\;qubits}\\
 \lstick{\ket{b_3}} & \qw  & \qw & \qw & \qw & \qswap \qw & \qw  & \qw  \gategroup{4}{7}{5}{8}{1em}{\}}  \\
}
\]
\caption[The controlled swap of two pairs of two qubits]{Using controlled quantum Swap gates to exchange two pairs of two qubits. Note the use of quantum superposition to build the quantum state.}
\label{fig15}
\end{center}
\end{figure}
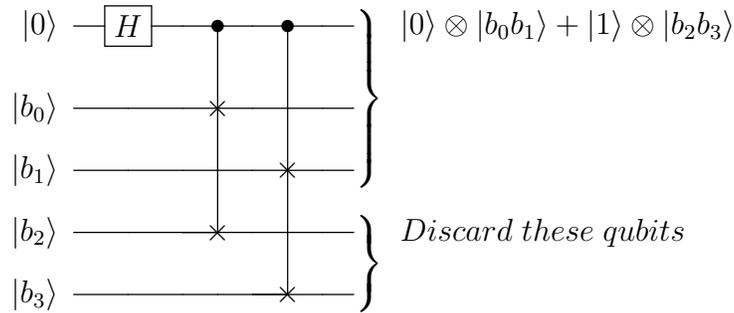

The controlled swap of multiple pairs of qubits is used extensively in the quantum circuits to follow. A quantum circuit implementing the controlled swap of two pairs of qubits is shown in Figure \ref{fig15}.

\subsection{Reusing {\em Discarded} qubits}
\label{reusequbits}

\begin{figure}[h!]
\begin{center}
\[
\Qcircuit @C=1.0em @R=2em {
\lstick{|q_1\rangle\,=\;|0\rangle} & \gate{H} &\qw &  \ctrl{1} & \qw & \qw & \qw& \rstick{ |0\rangle \otimes |b_0\rangle  + |1\rangle \otimes |b_1\rangle } \\
 \lstick{|q_2\rangle\,=\,\ket{b_0}} & \qw  & \qw & \qswap \qwx[1] & \qw & \qw  & \qw   \gategroup{1}{6}{2}{7}{1em}{\}}\\
  \lstick{|q_3\rangle\,=\,\ket{b_1}} & \qw  & \qw & \qswap \qw & \qw  & \qw &\qw & \rstick{Discard\;this\;qubit}\\
}
\]
\caption{The Controlled Swap (CSwap) gate.}
\label{fig14b}
\end{center}
\end{figure}
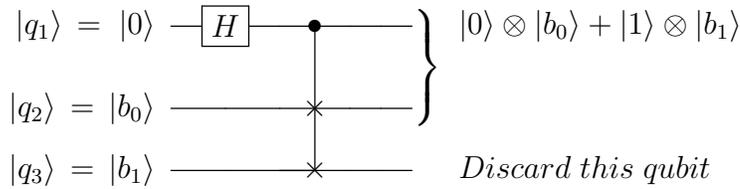

The quantum circuits shown in Figures \ref{figa}, \ref{fig14}, \ref{figh}, \ref{figi} and \ref{fig15} all
have {\em Discarded} 
qubits at the end of the computation. Ideally one would like to reuse the {\em Discarded} qubits later in the computation. However the {\em Discarded} qubits are typically entangled with other qubits in the circuit, complicating the reuse of the discarded qubits in other circuit blocks. The qubit reuse problem is shown in Figure \ref{fig14b}, which is the same circuit as Figure \ref{fig14} on page \pageref{fig14} but with 
the qubit labels $|\,q_1,\,q_2,\,q_3\,\rangle$ added for clarity. 

In Figure \ref{fig14b}, the variables $b_0$ and $b_1$ are individually 
either $0$ or $1$. There are four (4) possible 
$\{ \,b_0,\,b_1\,\}$ pairings. 
In Figure \ref{fig14b} the final circuit state is

\begin{equation}
|\,q_1,\,q_2,\,q_3\,\rangle  \;=\; |\,0 \,b_0 \,b_1\,\rangle  \;+\; |\,1 \,b_1\, b_0\,\rangle  .
\label{ediscard1}
\end{equation}

Ideally, one would like to see the final circuit quantum state be as shown in Equation \ref{ediscard1b}.

\begin{equation}
|\,q_1,\,q_2,\,q_3\,\rangle  \;=\; \underbrace{\Big ( \; |\,0 \,b_0 \,\rangle  \;+\; |\,1 \,b_1\,\rangle \;\Big ) }_{\displaystyle |q_1,q_2\rangle}
\;\otimes \;\underbrace{|junk\rangle}_{\displaystyle |q_3\rangle}.
\label{ediscard1b}
\end{equation}

Given the single qubit {\em junk} state $|junk\rangle$ in Equation \ref{ediscard1b} 
is in a tensor product with the remaining two qubit state $|q_1,q_2\rangle$, the
{\em junk} state can be removed without affecting the $|q_1,q_2\rangle$ state. The {\em junk} state can
be reused as an ancilla or in some other role later in the computation without fear of impacting earlier completed computational operations. The removal and reuse of the {\em junk} qubit can occur even if there are classical correlations between the {\em junk} state $|junk\rangle$ and the state $|q_1,q_2\rangle$. While
entanglement between $|junk\rangle$ and $|q_1,q_2\rangle$ would allow a manipulation of  $|junk\rangle$ to affect the state of $|q_1,q_2\rangle$, correlations do not. 

Returning to the quantum state in Equation \ref{ediscard1} which is 
output by the circuit in Figure \ref{fig14b}, 
if $b_0 \, \neq \,b_1$, measuring the {\em Discarded} qubit $q_3$ would influence the remaining $|q_1\,q_2\rangle $ possibly superposition state. 
For example, if $b_0 \;=\; 0$ and $b_1\;=\; 1$, the state shown in Equation \ref{ediscard1} becomes

\begin{equation}
|\,q_1\,q_2\,q_3\,\rangle  \;=\; |\,0 \,0 \,1\,\rangle  \;+\; |\,1 \,1\, 0\,\rangle  .
\label{ediscard2}
\end{equation}

In this situation, 
obtaining a $q_3$ measurement outcome of $0$ would leave $| \,q_1 \,q_2\,\rangle \;=\; |\,1\,1\,\rangle$, which is not the desired 
$| \,q_1 \,q_2\,\rangle$ state of $| \,q_1 \,q_2\,\rangle \;=  |\,0 \,0\,\rangle  \;+\; |\,1 \,1\,\rangle$.
Similarly, obtaining a $q_3$ measurement outcome of $1$ would leave $| \,q_1 \,q_2\,\rangle \;= |\,0\,0\,\rangle$, which again is not the desired $| \,q_1 \,q_2\,\rangle$ state of $| \,q_1 \,q_2\,\rangle \;=  |\,0 \,0\,\rangle  \;+\; |\,1 \,1\,\rangle$.

For the choice of $b_0 \;=\; 0$ and $b_1\;=\; 1$ as in the measurement examples above, the {\em desired} form of $|\,q_1 \,q_2\,\rangle $ after a measurement 
of $q_3$ would leave $|\,q_1\,q_2\,\rangle$ in the state $ |\,q_1\,q_2\,\rangle  \;=\; |\,0 \,0 \,\rangle  \;+\; |\,1 \,1\,\rangle$.
The qubit reuse 
scenario requires post-processing of qubit $q_3$ so that measuring $q_3$ will not impact the
remaining $|\,q_1 \,q_2\,\rangle $ superposition state, leaving $|\,q_1\,q_2\,\rangle $ in the 
state $|\,q_1\,q_2\,\rangle  \;=\; |\,0 \,b_0 \,\rangle  \;+\; |\,1 \,b_1\,\rangle$ regardless of the measurement outcome 
of $q_3$. Mathematically one would like a disentangling processing $\mathcal{F}$ producing a three qubit state as shown in 
Equation \ref{ediscard3}. 

\begin{equation}
|\,q_1\,q_2\,q_3\,\rangle  \;\stackrel{\mathcal{F}}{\Longrightarrow} \; 
|\,q_1\,q_2\,\rangle \;\otimes \; | \,\widetilde{q}_3\,\rangle , 
\label{ediscard3}
\end{equation}

where $\widetilde{q}_3$ represents the state of the third qubit after the disentangling process and 
$|\,q_1\,q_2\,\rangle$ represents the state $|\,q_1\,q_2\,q_3\,\rangle$  with the state of $q_3$ traced out. 
The disentangling strategy allows {\em Discarded} qubits to be reused in 
the computation, lowering the overall number of qubits needed to implement 
the circuits to be discussed. 

\begin{figure}[h!]
\begin{center}
\[
\Qcircuit @C=1.0em @R=2em {
 & & & & & & & \mbox{{\em Disentangling}} & & & & & & & \\
 & & & & & & & \mbox{{\em Operation $\mathcal{F}$}} & & & & & & & \\
\lstick{|q_1\rangle\,=\, |0\rangle} & \gate{H} &\qw &  \ctrl{1} & \qw & \qw & \qw &  \qw & \qw & \ctrl{2} & \qw & \qw &  \qw & \qw& \rstick{ |0\rangle \otimes |b_0\rangle  + |1\rangle \otimes |b_1\rangle } \\
 \lstick{|q_2\rangle\,=\,\ket{b_0}} & \qw  & \qw & \qswap \qwx[1] & \qw & \ctrl{2}  &  \qw & \qw & \qw  & \qw & \qw & \qw &  \qw & \qw \gategroup{3}{10}{4}{14}{1em}{\}}\\
  \lstick{|q_3\rangle\,=\,\ket{b_1}} & \qw  & \qw & \qswap \qw & \qw  & \qw & \qw & \ctrl{1} &  \qw & \targ & \qw   & \qw &  \qw & \qw& \rstick{Discard\;this\;qubit}\\
\lstick{\ket{a}\,\equiv\, \ket{0}} & \qw  & \qw & \qw \qw & \qw & \targ  & \qw &\targ & \qw &  \ctrl{-1}  & \qw  & \qw &  \qw & \qw & \rstick{Discard \;this\;qubit} \gategroup{3}{6}{6}{10}{1em}{--}\\
}
\]
\caption[The Controlled Swap gate with an ancilla used to remove entanglement]{The Controlled Swap gate with an ancilla $\ket{a}$ used to remove entanglement between qubits.}
\label{fig14withancilla}
\end{center}
\end{figure}
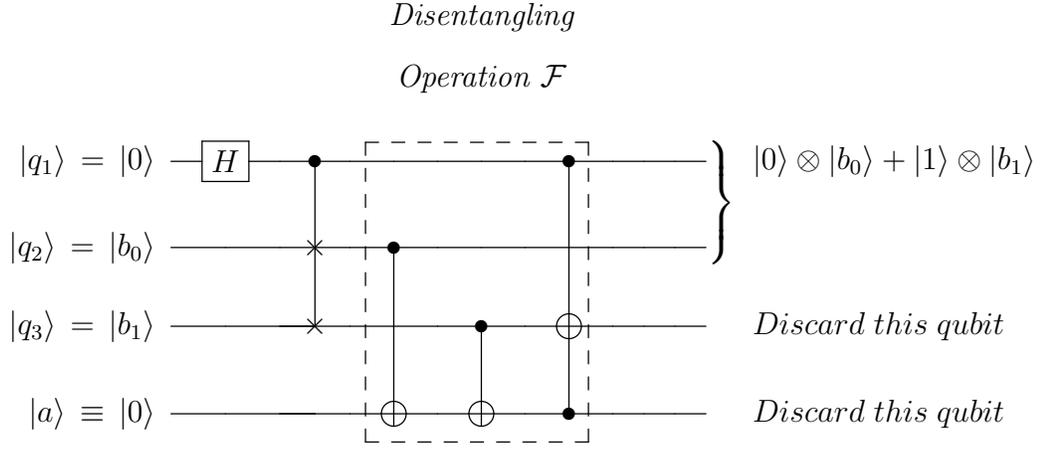

The inspiration for the disentangling approach is taken from Quantum Error Correction (QEC) techniques. 
In the circuit shown in Figure \ref{fig14withancilla}, the three gates within the dashed box implement the disentangling operation referred to above as $\mathcal{F}$,
which is a parity checking function as implemented for a variety of quantum codes. The innovation is the implementation
of error correction circuitry to enable qubit reuse in a computation.   
The three terminal Toffoli gate inside the dashed disentangling box $\mathcal{F}$ of Figure \ref{fig14withancilla}
can be decomposed into 
CNOT's and single qubit unitaries. 
Table \ref{table3} lists the quantum states at various points in the circuit shown in Figure \ref{fig14withancilla} for the 
four cases $b_0=\{0,1\}$ and $b_1=\{0,1\}$. The ancilla qubit $|a\rangle$ is always $|0\rangle$ at the input of the circuit. As in 
quantum error correction protocols, the ancilla qubit is measuring the parity of $b_0$ and $b_1$, which is equivalent to the sum
$a\;=\; b_0 \,\oplus \,b_1$ modulo 2.

Measuring the $|q_3\rangle$ qubit and the ancilla qubit $|a\rangle$ indicates which of the four bit pairs $\{b_0,b_1\}$ were 
loaded into the two qubit state $|\,q_1,\,q_2\,\rangle$. Unlike the situation in quantum error correction,
the measurement of 
$|q_3\rangle$ and the ancilla qubit $|a\rangle$ will not destroy the desired superposition of the remaining qubits. One already knows
what $b_0$ and $b_1$ are from the original classical bit specification, 
so the measurement does not gain any additional information about $b_0$ and $b_1$ from the $|q_3\rangle$ and the ancilla qubit $|a\rangle$ 
measurement outcomes. 

\begin{table}[h!]
\begin{tabular}{||c || c || c ||c|| c||}
\hline \hline 
 & &Input State& State $|q_1,  q_2, q_3, a\rangle$& \\
$b_0$ & $b_1$ & {\footnotesize $|q_1,  q_2, q_3, a\rangle$} & before $\mathcal{F}$ & State $|q_1,  q_2, q_3, a\rangle$ after $\mathcal{F}$\\
\hline \hline 
0 & 0 &$|\,0 \,0\,0 \,0\, \rangle$ & $ |\,0 \,0 \,0\,0\,\rangle  + |\,1 \,0\, 0\,0\,\rangle  $ & 
$ |\,0 \,0 \,0\,0\,\rangle  + |\,1 \,0\, 0\,0\, \rangle = $ {\boldmath$ \left (  |\,0 \,0 \,\rangle  +|\,1 \,0\, \rangle  \right )$} $ \otimes \, |\,0\, 0\,\rangle$ \\
\hline \hline 
0 & 1 &$|\,0 \,0\,1 \,0\, \rangle$ & $ |\,0 \,0 \,1\,0\,\rangle  + |\,1 \,1\,0\,0\,\rangle  $ & 
$ |\,0 \,0 \,1\,1\,\rangle  + |\,1 \,1\, 1\,1\, \rangle = $ {\boldmath$ \left (  |\,0 \,0 \,\rangle  + |\,1 \,1\, \rangle  \right )$} $ \otimes \, |\,1\, 1\,\rangle$ \\
\hline \hline 
1 & 0 &$|\,0 \,1\,0 \,0\, \rangle$ & $ |\,0 \,1 \,0\,0\,\rangle  + |\,1 \,0\, 1\,0\,\rangle  $ & 
$ |\,0 \,1 \,0\,1\,\rangle  + |\,1 \,0\, 0\,1\, \rangle =$ {\boldmath$ \left (  |\,0 \,1 \,\rangle  + |\,1 \,0\, \rangle  \right ) $} $ \otimes \, |\,0\, 1\,\rangle$ \\
\hline \hline 
1 & 1 &$|\,0 \,1\,1 \,0\, \rangle$ & $ |\,0 \,1 \,1\,0\,\rangle  + |\,1 \,1\, 1\,0\,\rangle  $ & 
$ |\,0 \,1 \,1\,0\,\rangle  + |\,1 \,1\, 1\,0\, \rangle = $ {\boldmath$\left (  |\,0 \,1 \,\rangle  + |\,1 \,1\, \rangle  \right )$} $ \otimes \, |\,1\, 0\,\rangle$ \\
\hline \hline  
\end{tabular}
\caption[Tracking qubit states in a qubit erasure circuit]{The four qubit state $|\,q_1\,q_2\,q_3\,a\, \rangle$ at three locations in the quantum circuit shown in Figure \ref{fig14withancilla} moving left to right. The final two qubit state $|\,q_1\,q_2\,\rangle$ is shown in {\bf bold} in the column labeled {\em State after $\mathcal{F}$}.} 
\label{table3}
\end{table} 

The qubits $|\,q_3\,\rangle$ and $|\,a\,\rangle$ can be measured or left untouched at the end of the disentangling circuit. In either case these qubits, either singly or together, can be used in later parts of the computation. The $|\,q_3\,\rangle$ and $|\,a\,\rangle$ qubits have been stripped by the disentangling operation $\mathcal{F}$ of their entangled connection to the $|\,q_1 \,q_2\,\rangle$ qubit state.

The same approach can be used iteratively to remove multiple {\em discarded qubits} in a circuit. Referring 
to Figure \ref{figa}, one can apply the disentangling method to decouple all four (4) of the discarded qubits and make these four qubits  available 
to computational subcircuits downstream without influencing the three qubit superposition state $|\,\alpha\,\beta\,\gamma\,\rangle$.\footnote{Think of the iterative disentangling application as peeling an onion, with each individual peeling action the application of the disentangling operation to one qubit.} 

\subsubsection{Optimizing the data flow from classical bits into quantum states}

\begin{figure}[h!]
\centering
\tikzstyle{block1} = [rectangle, draw, fill=red!30, text centered,rounded corners]
\tikzstyle{block2} = [rectangle, draw, fill=green!30, text centered,rounded corners]
\begin{tikzpicture}[node distance=5cm]
\node [block1,text width=4cm] (start) {Classical Input Data $\{\, b_0, b_1, \cdots, b_N\,\}$ };
\node [block2,right of =start,text width=3cm] (rawdata) {Quantum State $|\, b_0 \,b_1 \,\cdots \,b_N\,\rangle$ };
\path[line width=3pt,red] (start) edge[->] (rawdata);
\end{tikzpicture}
\caption[Classical to quantum data flow]{Classical to quantum data flow. The Red shaded boxes denote classical domain data. Green shaded boxes denote quantum data. The output quantum state is a product state: $|\, b_0 \,b_1 \,\cdots \,b_N\,\rangle\;=\; |\, b_0\,\rangle \, \otimes \, |\, b_1\,\rangle \, \otimes \,\cdots \,   \otimes\,|\,
b_N\,\rangle$.}
\label{fit1}
\end{figure}
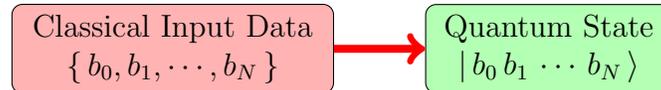

Information theory provides guidance on how to think about optimally
transferring bits from the classical domain to the quantum domain
using the minimum number of qubits and gates in the transfer operation. A block diagram of the transfer process is given in Figure \ref{fit1}. 

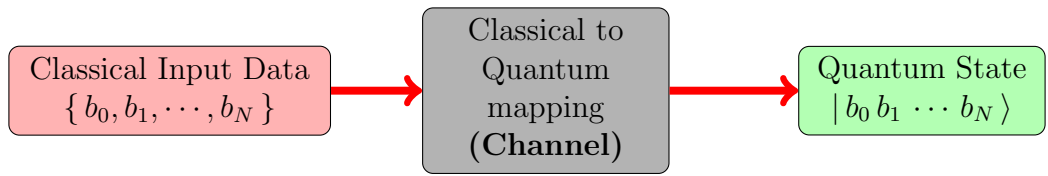
\begin{figure}[h!]
\centering
\tikzstyle{block1} = [rectangle, draw, fill=red!30, text centered,rounded corners]
\tikzstyle{block2} = [rectangle, draw, fill=green!30, text centered,rounded corners]
\tikzstyle{block3} = [rectangle, draw, fill=black!30, text centered,rounded corners]
\begin{tikzpicture}[node distance=5cm]
\node [block1,text width=4cm] (start) {Classical Input Data $\{\, b_0, b_1, \cdots, b_N\,\}$ };
\node [block3,right of =start,text width=3cm] (channel) {Classical to Quantum mapping {\bf (Channel)}};
\node [block2,right of =channel,text width=3cm] (rawdata) {Quantum State $|\, b_0 \,b_1 \,\cdots \,b_N\,\rangle$ };
\path[line width=3pt,red] (start) edge[->] (channel);
\path[line width=3pt,red] (channel) edge[->] (rawdata);
\end{tikzpicture}
\caption[Data flow through a classical to quantum channel]{Classical to quantum data flow through a quantum channel. The Red shaded boxes denote classical domain data. Green shaded boxes denote quantum data. The black box will use the circuits shown in Figures \ref{figb} and \ref{fige}.}
\label{fit2}
\end{figure}

The process of transferring bits from the classical domain to the quantum domain is a {\em channel} in information theoretic language. 
The picture of the data transfer dynamics is represented as shown in Figure \ref{fit2}. 

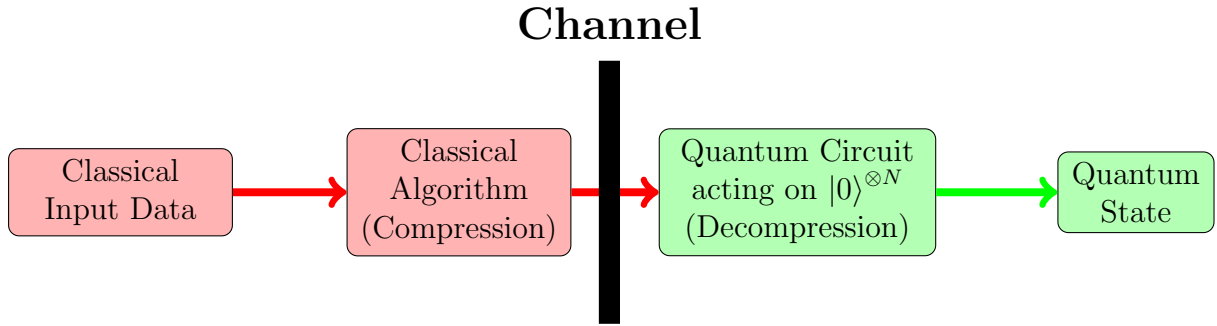
\begin{figure}[h!]
\tikzstyle{block1} = [rectangle, draw, fill=red!30, text centered,rounded corners]
\tikzstyle{block2} = [rectangle, draw, fill=green!30, text centered,rounded corners]
\begin{tikzpicture}[node distance=4.5cm]
\node [block1,text width=2.7cm] (start) {Classical Input Data};
\node [block1,right of =start,text width=2.7cm] (rawdata) {Classical Algorithm (Compression)};
\node [block2, right of =rawdata,text width=3.4cm] (quantumcircuit) {Quantum Circuit acting on $\left |0\right \rangle^{\otimes N}$ (Decompression)};
\path[line width=3pt,red] (start) edge[->] (rawdata);
\path[line width=3pt,red] (rawdata) edge[->] (quantumcircuit);
\node [block2, right of =quantumcircuit,text width=1.8cm] (quantumstate) {Quantum State};
\path[line width=3pt,green] (quantumcircuit) edge[->] (quantumstate);
\draw[line width=8pt,black] (6.5,-1.75) -- (6.5,1.75) ;
\node at (6.5,2.25) {\bf {\Large Channel}};
\end{tikzpicture}
\caption[Optimizing data flow by using compression techniques]{To optimize data flow, compress binary data in the classical domain and decompress in the quantum domain.}
\label{fit3}
\end{figure}

Claude Shannon's channel capacity theorem, together with the
source coding theorem, indicate that to optimize the classical to quantum transfer of data, one should compress the incoming bit stream to remove any redundancies among the input bits, transmit the compressed data through the channel, and decompress the data stream at the channel output to recover the original data bits.  This sequence of operations is shown in Figure \ref{fit3}. 

Any improvement in data transfer rate depends on the 
compressibility of the incoming data stream. For incoming bits which are 
statistically {\bf I}ndependent, {\bf I}dentically {\bf D}istributed ({\bf IID}) and 
equiprobable, there is no benefit to using a compression/decompression scheme. 
However, in other scenarios there will be a benefit. 

\begin{figure}[h!]
\tikzstyle{block1} = [rectangle, draw, fill=red!30, text centered,rounded corners]
\tikzstyle{block2} = [rectangle, draw, fill=green!30, text centered,rounded corners]
\begin{tikzpicture}[node distance=5cm]
\node [block1,text width=3.25cm] (start) {Raw Data blocks of length $N$ bits};
\node [block1,right of =start,text width=2.5cm] (rawdata) {Classical Compression};
\node [block1, right of =rawdata,text width=4cm] (compressedstrings) {Compressed blocks of length\\$M\,=\,N\,L\,\le \,N$ bits};
\path[line width=3pt] (start) edge[->] (rawdata);
\path[line width=3pt] (rawdata) edge[->] (compressedstrings);
\end{tikzpicture}
\caption[Non-trivial compression block diagram]{The compression rate is $\frac{1}{\;L\;}$, with $ L\, >\, 1$. The red 
shading indicates all three blocks are implemented in the classical domain.}
\label{figsparse3}
\end{figure}
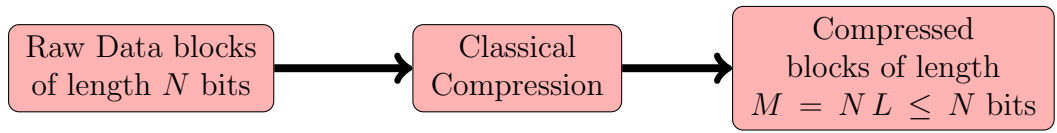

\begin{figure}[h!]
\tikzstyle{block1} = [rectangle, draw, fill=red!30, text centered,rounded corners]
\tikzstyle{block2} = [rectangle, draw, fill=green!30, text centered,rounded corners]
\begin{tikzpicture}[node distance=5.5cm]
\node [block1,text width=3.25cm] (start) {Raw Data blocks of length $N$ bits};
\node [block1, right of =start,text width=4cm] (compressedstrings) {Compressed blocks of length\\$M=NL\,\le \,N$ bits};
\node [block2, right of =compressedstrings,text width=4cm] (qcompressedstrings) {Blocks of length\\$M=NL\,\le \,N$ qubits};
\path[line width=3pt] (start) edge[->] (compressedstrings);
\path[line width=3pt] (compressedstrings) edge[->] (qcompressedstrings);
\end{tikzpicture}
\caption[Quantifying circuit complexity reduction using compression techniques]{Quantifying circuit complexity reduction using compression techniques. The circuits shown in Figure \ref{fige} become $M$ parallel stages instead of the $N$ parallel stages shown. 
The number of qubits saved by the compression/decompression  approach is $N\,-\,M$. The red shaded blocks occur
in the classical domain, while the green block is in the quantum domain.}
\label{figsparse4}
\end{figure}
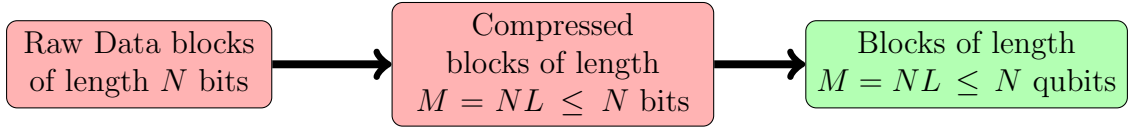

Figures \ref{figsparse3} and \ref{figsparse4} present the basic concepts of 
the compression/decompression approach. In Figure \ref{figsparse3},  a block of $N$ bits is compressed by a factor of 
$L$ to a block of $M$ bits, where $M \le N$. 
The factor $L\;=\;-\,E_\mathcal{S}\left[\,Log\left(\,p_i\,\right)\,\right]$ is the average entropy of a bit 
in the incoming bit stream. For large blocks, meaning $N >> 1$, a single bit is mapped to $L$ bits, where $0 \le L \le 1$. Therefore the block of length $N$ bits is mapped by the compression algorithm to $L\,N\;=\;M$ bits, where $0\le M \le N$. 
In our scheme the compression algorithm is implemented on a classical computer. 

\vspace*{0.3in}

\begin{figure}[h!]
\includegraphics[width=7.5in,height=4in]{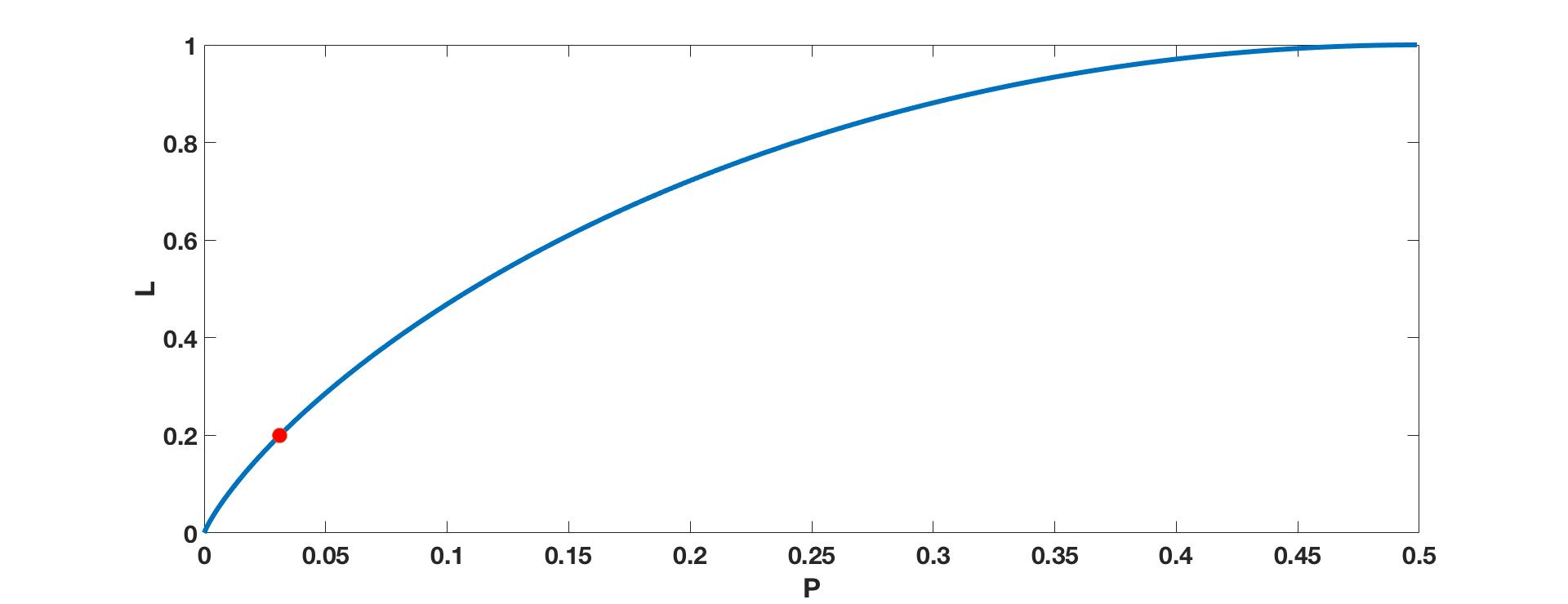}
\caption[Quantifying circuit complexity reduction using compression techniques II]{The circuits in Figure \ref{fige} become $M$ parallel stages instead of the $N$ parallel stages shown in Figure \ref{fige}. 
The number of qubits saved by the use of the compression/decompression  approach is $N\,-\,M$. The \textcolor{red}{red dot} indicates the $(\,p\,=\,0.03,\;L\,=\,0.2\,)$ operating point discussed in the main body.}
\label{figsparse5}
\end{figure}

The improvement in data transfer performance can be quantified in terms of the number of bit to qubit stages such as shown in Figure \ref{fige} on page \pageref{fige} needed to transfer $N$ bits of information. For example, for IID classical bits for which binary 1's occur with probability $p$ and 0's occur with probability $1-p$, with
$ 0\, <\, p \,\ll\, 1-p \,< \,1$, the compression parameter $L$ defined above 
behaves as shown in Figure \ref{figsparse5}. Working from the curve in Figure \ref{figsparse5}, when $p=0.03$, then
$M=0.2 N \;=\; \frac{N}{5}$. ( For this operating point, please see the \textcolor{red}{red dot} on the curve in Figure \ref{figsparse5}. ) 
When only $3\%$ of the incoming bits are on average 1's, compression/decompression methods can reduce the number of
loading circuits shown in Figure \ref{fige} from N to $\frac{N}{5}$. For $N=100$ and $p=0.03$, the number of stages needed in the circuit shown in Figure \ref{fige} is $20$, yielding a savings in the number of front end qubits and gate count through the use of compression/decompression techniques 
of $N\,-\,M\;=\;80$ qubits. This should be compared to approaches which do not use compression/decompression methods,
which for the 
circuit shown in Figure \ref{fige} would require $N=100$ qubit stages to transfer $100$ uncompressed classical bits. 

\subsubsection{Computational Complexity of Compression and Decompression}

Referring to 
the leftmost (Green) block in
Figure \ref{fig3} on page \pageref{fig3}, to ensure 
the data loading circuit implementation is a time complexity of order $\mathcal{O}\left [ \,Log(N)\,\right ]$, 
the quantum gate circuit depth of the decompression stage must be of order $\mathcal{O}\left [ \,Log(N)\,\right ]$. 

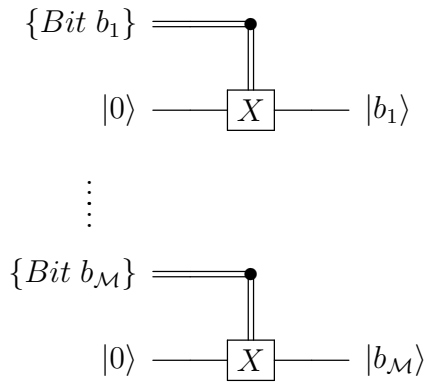
\begin{figure}[h!]
\begin{center}
\[
\Qcircuit @C=1.2em @R=2.0em {
 \lstick{\{Bit \;b_1\} } & \cw & \control \cwx[1] \cw \\
\lstick{|0\rangle} & \qw & \gate{X}  & \qw & \rstick{|b_1\rangle} \qw & & & & & & & & \\
}
\]
$$
\vdots \qquad \qquad \qquad  \qquad \qquad \qquad  \qquad \qquad \qquad  \qquad
$$
\vspace*{-0.4in}
$$
\vdots \qquad \qquad \qquad  \qquad \qquad \qquad  \qquad \qquad \qquad  \qquad
$$
\[
\Qcircuit @C=1.2em @R=2.0em {
 \lstick{\{Bit \;b_\mathcal{M}\} } & \cw & \control \cwx[1] \cw \\
\lstick{|0\rangle} & \qw & \gate{X}  & \qw & \rstick{|b_\mathcal{M}\rangle} \qw & & & & & & & & \\
}
\]
\caption[Loading $\mathcal{M}$ bits into $\mathcal{M}$ qubits in a single gate depth quantum circuit]{Loading $\mathcal{M}$ classical bits $\{ b_1,\cdots,b_\mathcal{M}\}$ into $\mathcal{M}$ qubits in a quantum circuit with a gate depth equal to one.}
\label{figeb}
\end{center}
\end{figure}

A proof that the $\mathcal{O}\left [ \,Log(N)\,\right ]$ decompression bound is achievable proceeds as follows. 
Recall that although
the data structures involved in the quantum decompression algorithm are qubits, the qubits are storing classical bit values and are in the pure state $|0\rangle$ or $|1\rangle$ at the beginning of the decompression circuit. 
This fact is due to the nature of the data loading circuitry shown in Figures \ref{figeb} which serves to transfer 
the $M$ compressed bits output from the classical compression algorithm from the classical domain into the quantum domain. 

As the qubits $|b_k\rangle$, $k=1,\cdots,M$, are individually 
either $|0\rangle$ or $|1\rangle$, with no quantum superposition present, the orthogonality of the pure states
single qubit $|0\rangle$ or $|1\rangle$ allows the quantum decompression algorithm and 
corresponding quantum decompression circuit 
to be a quantum version of the classical decompression corresponding to the classical compression algorithm which was used. 
The classical gates in the classical decompression circuit (e.g. NAND) are implemented using quantum gates. 
In this manner, the extensive literature of classical decompression algorithms and circuits 
can be leveraged. 

Using the fact that one can represent each classical gate in the classical decompression algorithm
with a quantum gate configuration consisting of, at most, a fixed, finite number of quantum gates, the proof of $\mathcal{O}\left [ \,Log(N)\,\right ]$ quantum gate depth for the quantum decompression circuit follows from the corresponding 
proof of $\mathcal{O}\left [ \,Log(N)\,\right ]$ gate depth of the classical 
decompression circuit. As there are many classical decompression algorithms with 
$\mathcal{O}\left [ \,Log(N)\,\right ]$ classical gate depth,
this concludes the discussion of the application of classical compression/decompression algorithms to
optimize the transfer of classical bits into qubits. 

\section{The recursive nature of Circuit Family \#2}

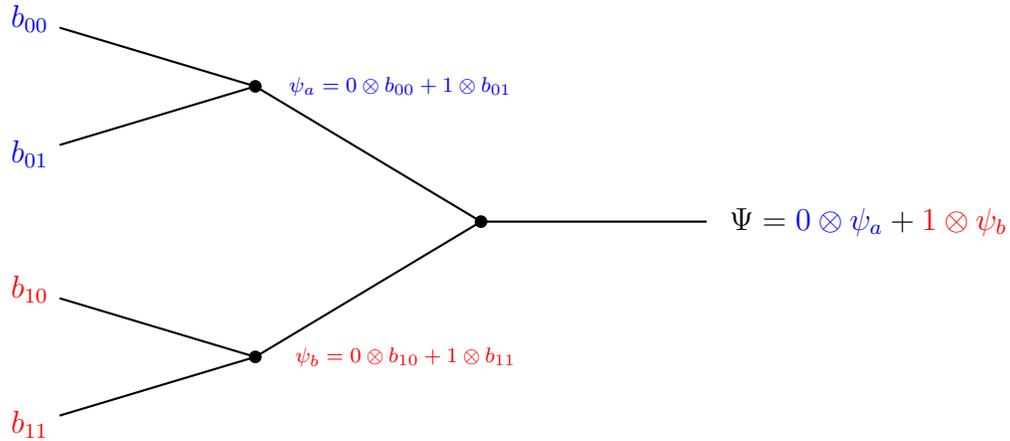
\begin{figure}[h!]
\begin{center}
\begin{tikzpicture}
[grow=-180,level distance=30mm,thick,
level 1/.style={sibling distance=72mm},
level 2/.style={sibling distance=36mm},
level 3/.style={sibling distance=18mm},
]
\coordinate
child { [fill] circle(2pt)
child {  [fill] circle(2pt) child {node {\textcolor{blue}{$b_{00}$}}}  child {node {\textcolor{blue}{$b_{01}$}}} {node {\hspace*{1.4in}{\textcolor{blue}{\scriptsize \quad $\psi_a=0\otimes b_{00}+1\otimes b_{01}$}}}}}
child { [fill] circle(2pt)  child {node {\textcolor{red}{$b_{10}$}}} child {node {\textcolor{red}{$b_{11}$}}} 
 {node {\hspace*{1.4in} {\textcolor{red}{\scriptsize \quad $\psi_b=0\otimes b_{10}+1\otimes b_{11}$}}}}}
{node {\hspace*{4in} $\Psi=\textcolor{blue}{0\otimes \psi_a} + \textcolor{red}{1\otimes \psi_b}$}}
};
\end{tikzpicture}
\caption[The recursive assembly of the quantum state $\Psi$]{The recursive assembly of the quantum state $\Psi\;=\; |00b_{00}\rangle \,+\, |01b_{01} \rangle \,+\, |10b_{10}\rangle  \,+\, |11b_{11}\rangle$. The depth of the recursion is $Log_2(\mathcal{N}) \,=\, Log_2(4) \,= \,2$.}
\label{figj}
\end{center}
\end{figure}

The circuit shown in Figure \ref{figa} on page \pageref{figa} has a recursive structure which enables the 
asymptotic behavior of circuit family \#2 to be extrapolated for large $\mathcal{N}$. 
The recursive nature of the assembly of the quantum state 

\begin{equation}
\psi_A\;=\;|0 0\rangle \otimes |b_{00}\rangle\, +\,|01\rangle \otimes |b_{01}\rangle \, +\, |10\rangle \otimes | b_{10}\rangle\, +\,|11\rangle\otimes |b_{11}\rangle 
\end{equation}
 
for the $2$ by $2$ matrix 
$A\;=\; \bmatrix{b_{00} \quad b_{01} \cr b_{10} \quad b_{11} }$
is shown in Figure \ref{figj}. Recall that each entry $b_{ij}$ in the matrix $A$ is a single classical bit. 

Consideration of the recursive tree construction in Figure \ref{figj} allows for the calculation of the number of qubits needed, as well as the gate depth for assembling, a quantum state containing $\mathcal{N}$ classical bits. 

\subsection{Resource tabulation for Circuit Family \#2}

The general approach to loading $\mathcal{N}$ classical bits into $Log_2(\mathcal{N})$ qubits is a generalization of the recursive method shown in Figure \ref{figj} for $\mathcal{N}=2^2$. 
Without loss of generality for asymptotic calculations, let $\mathcal{N}$ be a power of 2 and define $n$ such that $\mathcal{N}=2^n$. In this case there will be a total of $n$ levels of recursion in the quantum circuit 
loading $\mathcal{N}$ classical bits into $Log_2(\mathcal{N})$ qubits. 

\subsubsection[Asymptotic Quantum Gate Depth Calculation for Circuit Family \#2]{Asymptotic Quantum Gate Depth and Execution Time Resource Calculation for Circuit Family \#2}
 
The data loading circuit depth must obey bounds compatible with the quantum algorithm to be implemented. For an exponential speedup over a classical polynomial complexity algorithm, the data loading circuit depth must 
scale ideally as $\mathcal{O}\Big [\,Log\Big (\,\mathcal{N}\, \Big)\,\Big]$ and at most polylogarithmically in 
$Log\Big (\,\mathcal{N}\, \Big )$.\footnote{See Figure \ref{fig3}.} For circuit family \#2, the gate depth can be computed with the aid of the recursive tree diagram in Figure \ref{figj}, together with the controlled swap based circuits shown in Figures \ref{figa}, \ref{fig14} and \ref{fig15}. Moving from left to right in Figure \ref{figj}, the first layer requires one Hadamard gate and one controlled swap gate (CSwap). The second layer requires one Hadamard and two CSwap gates. The third layer requires one Hadamard and three
CSwap gates. And so on. Tabulating the number of gates in these recursion layers leads to the summations in Equations \ref{eqn100} and \ref{eqn101}.

\begin{equation}
\# \quad of \quad Hadamards\quad = \quad n \quad = \quad Log(\mathcal{N}).
\label{eqn100}
\end{equation}

\begin{equation}
\# \quad of\quad CSwap's\quad = \quad\sum_{k=1}^{k=n}\quad k \quad=\quad \frac{n(n+1)}{2} \quad \equiv \quad\mathcal{O}(n^2). 
\label{eqn101}
\end{equation}

Tallying all the gates in a generic family \#2 circuit from start to finish yields the total gate depth. 

\begin{equation}
\mathcal{O}(n)\;+\; \mathcal{O}(n^2)\;=\; \mathcal{O}(n^2)\; 
\equiv \;\mathcal{O}\Bigg [ \, \Big (Log_2\left \{\,\mathcal{N}\,\right \}\Big )^2\, \Bigg ].
\label{circuitfamily2gatedepth}
\end{equation}

The total gate depth is proportional to time. Therefore the asymptotic 
time scaling for circuit family \#2 is $\mathcal{O}\Bigg [ \Big (Log_2\big \{\,\mathcal{N}\,\big \}\Big )^2\Bigg ]$,
which is acceptable, but not ideal, for an exponential speedup quantum algorithm. Ideally one would like a gate depth which scales at most as 
$\mathcal{O}\Big [\,Log\Big (\,\mathcal{N}\, \Big)\,\Big]$.

\subsection{Asymptotic Space (Qubit) Resource Calculation\\for Circuit Family \#2}

Looking at Figure \ref{figj}, note the left most tree level uses $\mathcal{N}$ qubits and the next level to the right utilizes $\frac{\mathcal{N}}{2}$ {\it additional} qubits. Moving to the right, for generic $\mathcal{N}$, each succeeding level {\it additionally} needs a number of qubits equal to half of the number of qubits of the previous level.  There are a total number of $n\,=\,Log_2(\mathcal{N})$ levels. Tabulating the total number of qubits used by circuit family \#2 as a summation yields Equation \ref{eqn2}.

$$
Total \; \# \; of\;qubits \;=\; \sum_{k=1}^{k=n} \; \frac{\mathcal{N}}{2^{k-1}} \;=\; \mathcal{N} \; \Bigg ( \;1 \;+\; \frac{ \;1 \;}{2}\; +\; \frac{1}{ \;2^2 \;}\; +\; \frac{1}{ \;2^3 \;}\;+ \;\cdots\;+\; \frac{1}{ \;2^{n-1} \;}\ \; \Bigg ) 
$$

\begin{equation}
\;=\; \mathcal{N}\; \Bigg ( \;  \frac{\;1\;-\; 
\frac{1}{2^n}\;}{\;1\;-\; \frac{1}{2}\;}\;\Bigg ) \; =\; 2\,\mathcal{N} \; \Bigg ( \; 1\;-\; \frac{1}{\mathcal{N}}\;\Bigg )\;=\; 2\,\mathcal{N} \;-\; 2 \;< \; 2\,\mathcal{N}.
\label{eqn2}
\end{equation}

Note the use of the finite sum identity 

\begin{equation}
\sum_{k=0}^{k=n} \; x^k \;=\; 1\;+\;x \;+\;x^2 \;+\;\cdots \;+\; x^n \;=\; \frac{\;\;\;\;1\;-\; x^{n+1}\;}{1\;-\;x},
\end{equation} 

valid for $|x|< 1$, in Equations \ref{eqn2}, \ref{eHsum} and \ref{eCSsum1}. 

\renewcommand{\arraystretch}{1.25}
\begin{table}[h!]
\begin{center}
\begin{tabular}{||c | c | c |c| c|c||}
\hline \hline 
 & & Final & Final & Final &  \\
 &Number & number of & number & total &\\
 Circuit & of & qubits in& of & number & Gate \\
 family&classical&the quantum &ancilla& of  & depth \\
  &  bits  & state $\psi$&qubits (*)&qubits &\\
\hline 
\#1 & $\mathcal{N}$ & $\mathcal{N}$ &$0$ & $\mathcal{N}$ & $1$  \\
\hline
\#$2^{ne}$ & $\mathcal{N}$ & $\lceil \,Log_2(\mathcal{N})\,\rceil=n$ &$2\,\mathcal{N}-2-n$& 2$\mathcal{N}-2$&$\lceil \,Log_2(\mathcal{N})\,\rceil$ \\
\hline
\#$2^e$ & $\mathcal{N}$ & $\lceil \, Log_2(\mathcal{N})\,\rceil=n$ &$0$& $n$&$\lceil\, Log_2(\mathcal{N})\,\rceil$  \\
\hline \hline  
\end{tabular}
\end{center}
\caption[Resource requirements for data loading Circuit Families \#$1$ and \#$2$.]{Resource requirements for data loading circuit families \#$1$ and \#$2$. The total number of classical bits is $\mathcal{N}\;=\; 2^n$. Note that $\lceil$---$\rceil$ is the integer ceiling function. The circuit families $2^e$ and $2^{ne}$ stand for {\em erasure} and {\em no erasure}. The circuit family \#2 variations differ in whether ancilla qubits are erased and available for reuse upon completion of the data loading circuitry.
(* = Note that some or all of any remaining ancilla qubits at circuit completion may be entangled with the quantum state qubits.)}
\label{table2b}
\end{table}

\renewcommand{\arraystretch}{1.25}
\begin{table}[h!]
\begin{center}
\begin{tabular}{||c  |c| c|c|c| c|c||}
\hline \hline 
 &   Final & Final &  \multicolumn{4}{c||}{ }   \\
 &  number & total &\multicolumn{4}{|c||}{Total number} \\
 Circuit  &  of & number & \multicolumn{4}{|c||}{of gates} \\
 family&ancilla& of  &\multicolumn{4}{|c||}{ } \\
  &qubits (*)&qubits & \multicolumn{1}{c}{\fbox{CNOT}} & \multicolumn{1}{c}{\fbox{H}} & \multicolumn{1}{c}{\fbox{CSWAP}}& \multicolumn{1}{c||}{\fbox{CCNot}}\\
\hline 
\#1 & $0$ & $\mathcal{N}$ & $\mathcal{N}$ & $0$ &$0$ &$0$ \\
\hline
\#$2^{ne}$ &$2\,\mathcal{N}-2-n$& 2$\mathcal{N}-2$& $0$ & $\mathcal{N}-1$& $2\,\mathcal{N}\,-\,n\,-\,2$& $0$\\
\hline
\#$2^e$  &$0$& $n$& $2\;(2\,\mathcal{N}-2-n)$& $\mathcal{N}-1$&$2\,\mathcal{N}\,-\,n\,-\,2$& $2\,\mathcal{N}-2-n$\\
\hline \hline  
\end{tabular}
\end{center}
\caption[Updated resource requirements for data loading Circuit Families \#$1$ and \#$2$]{Resource requirements for data loading circuit families \#$1$ and \#$2$. The total number of classical bits is $\mathcal{N}\;=\; 2^n$. Note that $\lceil$---$\rceil$ is the integer ceiling function. The circuit families $2^e$ and $2^{ne}$ stand for {\em erasure} and {\em no erasure}. The erasure circuit family \#$2^e$ has additional circuitry to decouple discarded qubits from the circuit, erasing their contents and making the discarded qubits available for reuse. The circuit family \#2 variations \#$2^e$ versus \#$2^{ne}$ differ only in whether ancilla qubits are erased 
and available for reuse at circuit completion.
(* = Note that some or all of any remaining ancilla qubits at circuit completion may be entangled with the quantum state qubits.)}
\label{table2}
\end{table}

The time and space tabulation for circuit family \#2 added to the numbers in Table \ref{table1} on page \pageref{table1} yields Tables \ref{table2b} and \ref{table2}. Note that the tabulation of the total number of qubits for circuit family \#$2^{ne}$ does not permit ancilla qubit {\it reuse}. However for circuit family \#$2^e$, ancilla qubits are erased and these qubits are available for reuse in other portions of the circuit. In family \#$2^e$ qubits are reused as the master state $\psi$ 
is constructed, decreasing the overall spatial (qubit) resource requirements for the data loading circuit. 

\subsection{Computing total gate counts for Circuit Family \#2}

\subsubsection{The number of Hadamard gates}

Looking back to Figure \ref{figj} on page \pageref{figj} and the quantum circuits shown in Figures 
\ref{figa}, \ref{fig14} and \ref{fig15}, observe the overall tally of Hadamard gates is $\frac{\mathcal{N}}{2}$ for the first time slice, $\frac{\mathcal{N}}{4}$ for the second time slice, 
$\frac{\mathcal{N}}{8}$ for the third time slice, and so on. This summation is shown in Equation \ref{eHsum}.

\begin{equation}
\sum_{k=1}^{k=n}\; 
\frac{\mathcal{N}}{2^k}\;=\;  \frac{\mathcal{N}}{2} \; \sum_{k=0}^{k=n-1}\; \frac{1}{2^k} \;=\;   \frac{\mathcal{N}}{2} \; 
\frac{ 1-\frac{1}{2^n}}{1-\frac{1}{2}} \;=\; \mathcal{N}\;\left (\; 1\,-\, \frac{1}{\mathcal{N}}\;\right ) \;=\; \mathcal{N}-1,
\label{eHsum}
\end{equation}

which is entered under the column labelled $\fbox{H}$ in Table \ref{table2} for families \#$2^e$ and \#$2^{ne}$. 

\subsubsection{The number of Controlled Swap gates (CSwap's)}

Looking back to Figure \ref{figj} on page \pageref{figj} and the quantum circuits shown in Figures \ref{figa}, \ref{fig14} and \ref{fig15}, observe the tally of Controlled Swap (CSwap) gates is $\frac{\mathcal{N}}{2}$ for the first time slice, $2 \, \frac{\mathcal{N}}{4}$ for the second time slice, 
$3\, \frac{\mathcal{N}}{8}$ for the third time slice, and so on. This summation is shown in Equation \ref{eCSsum1}. 
Let $b$ be a variable which will later be set to $\frac{1}{2}$. 

\begin{equation}
\mathcal{N}\;
\sum_{k=1}^{k=n}\; 
\frac{k}{2^k}\;=\; \mathcal{N}\;
\sum_{k=1}^{k=n}\; 
k \,b^k\;=\; \mathcal{N}\;
b\; \frac{\partial}{\partial b} \; \sum_{k=1}^{k=n}\; 
b^k\;=\; 
\mathcal{N}\;
b\; \frac{\partial}{\partial b} \;\left ( \; \frac{1-b^{n+1}}{1-b}\, -\, 1\;\right ) 
\label{eCSsum1}
\end{equation}

\begin{equation}
=\; 
\mathcal{N}\;
b\; \left ( \; \frac{-\,(n+1) \,b^{n}}{1-b}\; -\; \frac{(1-b^{n+1})(-1)}{(1-b)^2}\;\right ) 
\;=\; 
\mathcal{N}\;\left ( \; -\,(n+1) \,b^{n}\; +\; 2\,(1-b^{n+1})\;\right ). 
\label{eCSsum2}
\end{equation}

As just mentioned, in Equation \ref{eCSsum2} set $b\,=\,\frac{1}{2}$ and 
note that $b^n\;=\; \frac{1}{\mathcal{N}}$, so $\mathcal{N}\,b^n\,=\,1$. Also note that $2b\,=\,1$. As a result Equation \ref{eCSsum2} becomes Equation \ref{eCSsum3}.

\begin{equation}
\mathcal{N}\;
\sum_{k=1}^{k=n}\; 
\frac{k}{2^k}\;=\;
 -\,(n+1) \; +\; 2\,\mathcal{N}\; -\;2\,b\,\mathcal{N} \,b^n\;=\;  -\,(n+1) \; +\; 2\,\mathcal{N}\; -\;1\;=\; 2\,\mathcal{N}\; -\; n \;-\; 2. 
\label{eCSsum3}
\end{equation}

The summation result shown in Equation \ref{eCSsum3} is entered under the column labelled $\fbox{CSWAP}$ in 
Table \ref{table2} for circuit families \#$2^e$ and \#$2^{ne}$. 
As a check, note that when $\mathcal{N}=4$, then $n\,=\,Log_2(\mathcal{N}) = 2$ and 
$ 2\,\mathcal{N}\; -\; n \;-\; 2 \;=\; 8 \,-\,2\,-\,2\;=\;4$. Referring to the $\mathcal{N}\,=\,4$ examples shown in
Figures \ref{figa} and \ref{figj}, on pages \pageref{figa} and \pageref{figj} respectively, 
one may verify that the number of CSwap's is indeed $4$. 

\subsubsection{The number of Controlled Not $\equiv$ CNOT gate's}

For circuit family \#$2^{ne}$, the number of controlled Not gates is zero. Looking back to the quantum erasure circuit in Figure \ref{fig14withancilla}, when the erasure circuitry of circuit 
family \#$2^e$ is included in the gate count, two controlled Not gates are used for every qubit discarded or erased. Since in circuit 
family \#$2^e$ the number of discarded and erased qubits is seen in Tables \ref{table2b} and \ref{table2} to be $2\,\mathcal{N}\;-\;2\;-\;n$, twice this number or $2\;(2\,\mathcal{N}\;-\;2\;-\;n\;)$ is entered in Table \ref{table2} under the Controlled Not gate count column. 

\subsubsection{The number of Toffoli gates $\equiv$ CCNot gate's}

For circuit family \#$2^{ne}$, the number of Toffoli gates is zero.\footnote{Do not count the Toffoli gate inside the Controlled Swap gate. That Toffoli gate is accounted for in the CSwap gate count.} Looking back to the quantum erasure 
circuit in Figure \ref{fig14withancilla}, when the erasure circuitry of circuit 
family \#$2^e$ is included in the gate count, one Toffoli is used for every qubit which is 
discarded or erased. Since in circuit 
family \#$2^e$ the number of discarded and erased qubits is seen in Tables \ref{table2b} and \ref{table2} to be $2\,\mathcal{N}\;-\;2\;-\;n$, this is the number of Toffoli gates entered in Table \ref{table2}. 

\section{Circuit Family \#3}
 
Circuit family \#$2$ presents the basic approach to data loading. In either of the two \#2 circuit families, $2^e$ or $2^{ne}$,
the
data loading circuit depth is the primary concern.\footnote{See Figure \ref{fig3}.} 
For circuit family \#2 the gate depth scaling with $\mathcal{N}$ 
was computed with the aid of the recursive tree diagram in Figure \ref{figj} and determined in Equation \ref{circuitfamily2gatedepth} on page \pageref{circuitfamily2gatedepth} to be 
$\mathcal{O}\Bigg [ \, \Big (Log_2\left \{\,\mathcal{N}\,\right \}\Big )^2\, \Bigg ].$ Ideally one would like a gate depth which scales at most as 
$\mathcal{O}\Big [\,Log\Big (\,\mathcal{N}\, \Big)\,\Big]$. 
Looking back at the gate depth computation for circuit family \#2, one finds 
the number and implementation of CSwap's is leading to the  $\mathcal{O}\Bigg [ \, \Big (Log_2\left \{\,\mathcal{N}\,\right \}\Big )^2\, \Bigg ]$ gate depth scaling behavior. Further examination of circuit family \#2's architecture
indicates the serial nature of the use of CSwap's in each time slice ultimately generates the limiting 
$\mathcal{O}\Big [\,\Big (\,Log\,\{\, \mathcal{N}\, \}\,\Big)^2\,\Big]$ scaling behavior. 

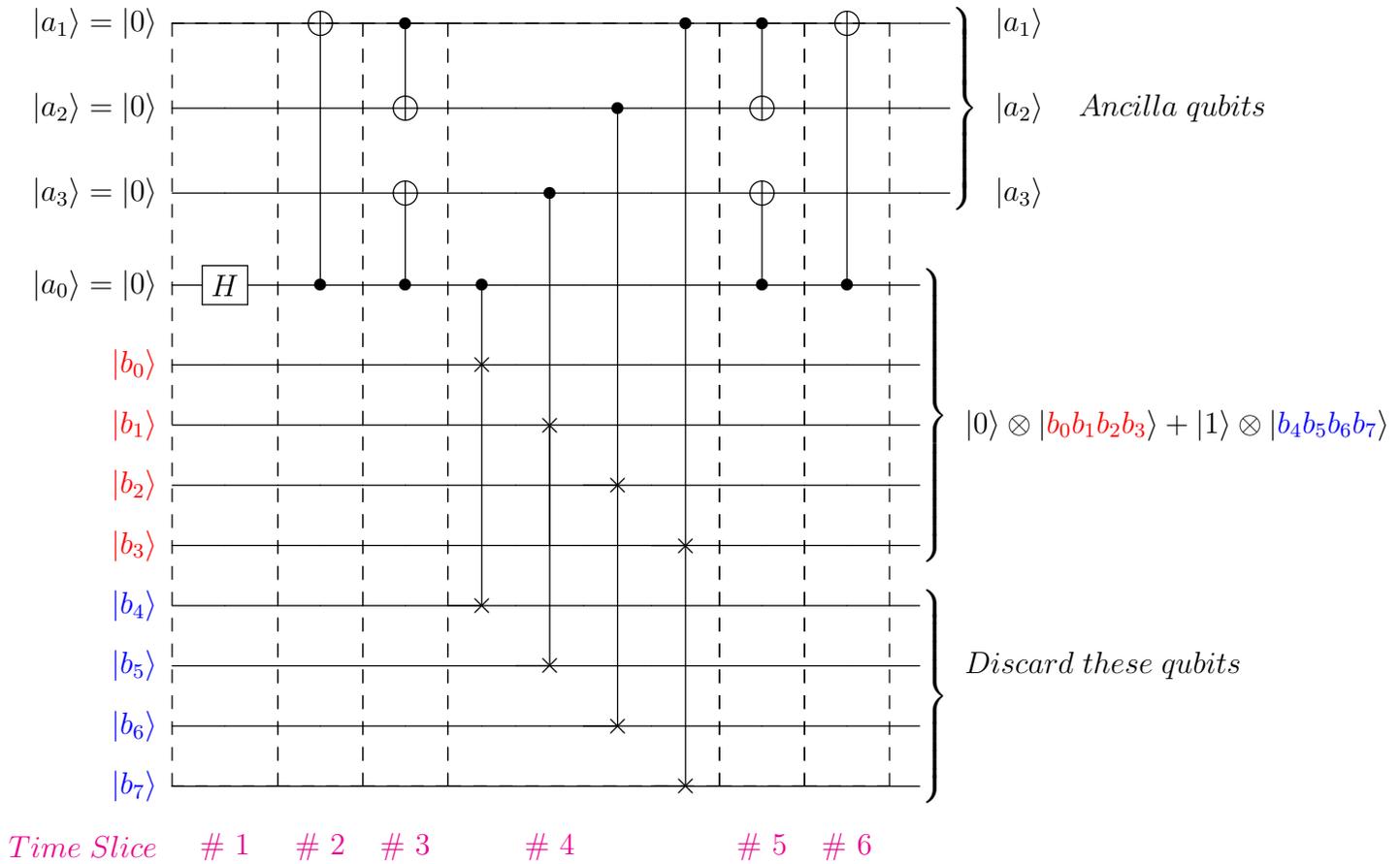
\begin{figure}[h!]
\begin{center}
\[
\hspace*{-0.5in}
\Qcircuit @C=1.0em @R=2em {
\lstick{\ket{a_1}=|0\rangle}  & \qw & \qw & \targ & \qw & \ctrl{1} & \qw & \qw & \qw & \qw & \qw & \qw & \qw   & \ctrl{11} & \qw & \ctrl{1} & \qw  & \targ & \qw & \qw & \qw & \rstick{\ket{a_1}} \gategroup{1}{20}{3}{21}{1em}{\}} \\
\lstick{\ket{a_2}=|0\rangle} & \qw & \qw & \qw & \qw & \targ & \qw & \qw & \qw & \qw & \qw  & \ctrl{9} & \qw & \qw & \qw  & \targ & \qw & \qw & \qw & \qw & \qw & \rstick{\ket{a_2}\quad Ancilla\;qubits} \gategroup{1}{1}{12}{3}{0em}{--} \gategroup{1}{3}{12}{5}{0em}{--}  \\
\lstick{\ket{a_3}=|0\rangle}  & \qw & \qw & \qw & \qw & \targ& \qw & \qw & \qw  & \ctrl{5} & \qw &  \qw & \qw & \qw & \qw & \targ & \qw & \qw & \qw & \qw & \qw & \rstick{\ket{a_3}} \gategroup{1}{5}{12}{7}{0em}{--} \gategroup{1}{7}{12}{15}{0em}{--}\\
\lstick{\ket{a_0}=|0\rangle} & \gate{H} &\qw & \ctrl{-3}    &\qw & \ctrl{-1} &  \qw & \ctrl{1} &\qw &  \qw & \qw & \qw & \qw& \qw & \qw & \ctrl{-1} & \qw  & \ctrl{-3} &\qw &\qw   \gategroup{1}{15}{12}{17}{0em}{--} \gategroup{1}{17}{12}{19}{0em}{--}\\
 \lstick{\textcolor{red}{\ket{b_0}}}  &\qw   & \qw &\qw &\qw   & \qw &\qw & \qswap \qwx[4] &\qw & \qw  &\qw & \qw  & \qw & \qw & \qw& \qw & \qw  &\qw &\qw &\qw \gategroup{4}{19}{8}{20}{1em}{\}}\\
 \lstick{\textcolor{red}{\ket{b_1}}} & \qw  &\qw &\qw &\qw &\qw & \qw &  \qw & \qw & \qswap \qwx[4]   & \qw & \qw & \qw& \qw & \qw& \qw & \qw  &\qw &\qw  &\qw & \rstick{ |0\rangle \otimes | \textcolor{red}{b_0b_1b_2b_3}\rangle  + |1\rangle \otimes |\textcolor{blue}{b_4b_5b_6b_7}\rangle }\\
 \lstick{\textcolor{red}{\ket{b_2}}} & \qw  &\qw &\qw &\qw &\qw & \qw & \qw & \qw & \qw & \qw  & \qswap \qw & \qw & \qw & \qw & \qw & \qw  &\qw &\qw &\qw \\ 
 \lstick{\textcolor{red}{\ket{b_3}}} & \qw  &\qw &\qw &\qw &\qw & \qw & \qw & \qw & \qw & \qw  & \qw & \qw & \qswap  \qw &\qw & \qw  & \qw & \qw& \qw  &\qw \gategroup{9}{19}{12}{20}{1em}{\}}  \\
 \lstick{\textcolor{blue}{\ket{b_4}}} & \qw  &\qw &\qw  & \qw  &\qw &\qw & \qswap \qw  & \qw &\qw   & \qw  & \qw & \qw & \qw& \qw & \qw & \qw  &\qw &\qw &\qw \\
 \lstick{\textcolor{blue}{\ket{b_5}}} & \qw  &\qw &\qw &\qw &\qw & \qw &  \qw &  \qw & \qswap \qw& \qw & \qw & \qw & \qw& \qw & \qw & \qw  &\qw &\qw &\qw & \rstick{Discard \;these\;qubits} \\
 \lstick{\textcolor{blue}{\ket{b_6}}} & \qw  &\qw &\qw &\qw &\qw & \qw &  \qw & \qw& \qw & \qw  & \qswap \qw & \qw & \qw & \qw& \qw & \qw  &\qw &\qw &\qw \\
 \lstick{\textcolor{blue}{\ket{b_7}}} & \qw  &\qw &\qw &\qw &\qw & \qw & \qw & \qw &  \qw & \qw  & \qw & \qw & \qswap  \qw & \qw& \qw  & \qw  &\qw &\qw &\qw  \\
 \lstick{\textcolor{magenta}{Time \;Slice}} & \mbox{\textcolor{magenta}{\# 1}}  & & \mbox{\textcolor{magenta}{\# 2}}  & & \mbox{\textcolor{magenta}{\# 3}}  & & & & \mbox{\textcolor{magenta}{\# 4}}  & & & & & & \mbox{\textcolor{magenta}{\# 5}} & &  \mbox{\textcolor{magenta}{\# 6}} &  & & & & \\
}
\]
\caption[Data loading Circuit family \#$3$ layout for $\mathcal{N}=8$]{Circuit family \#$3$ for $\mathcal{N}=8$. The controlled quantum swap gate for two pairs of four qubits or eight data qubits. Note the use of superposition to build the quantum state.}
\label{fig25}
\end{center}
\end{figure}

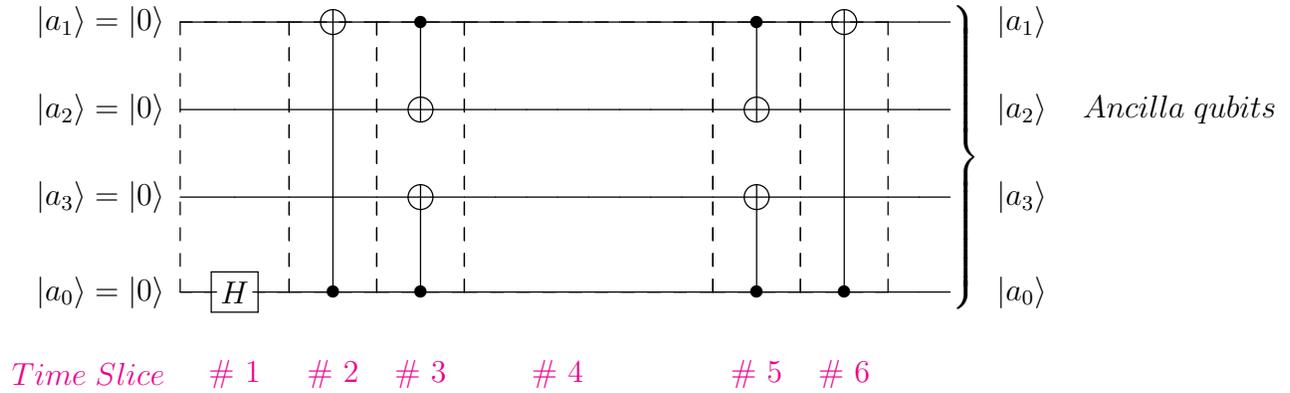
\begin{figure}[h!]
\begin{center}
\[
\hspace*{-0.5in}
\Qcircuit @C=1.0em @R=2em {
\lstick{\ket{a_1}=|0\rangle}  & \qw & \qw & \targ & \qw & \ctrl{1} & \qw & \qw & \qw & \qw & \qw & \qw & \qw   & \qw & \qw & \ctrl{1} & \qw  & \targ & \qw & \qw & \qw & \rstick{\ket{a_1}} \gategroup{1}{20}{4}{21}{1em}{\}} \\
\lstick{\ket{a_2}=|0\rangle} & \qw & \qw & \qw & \qw & \targ & \qw & \qw & \qw & \qw & \qw  & \qw & \qw & \qw & \qw  & \targ & \qw & \qw & \qw & \qw & \qw & \rstick{\ket{a_2}\quad Ancilla\;qubits} \gategroup{1}{1}{4}{3}{0em}{--} \gategroup{1}{3}{4}{5}{0em}{--}  \\
\lstick{\ket{a_3}=|0\rangle}  & \qw & \qw & \qw & \qw & \targ& \qw & \qw & \qw  & \qw & \qw &  \qw & \qw & \qw & \qw & \targ & \qw & \qw & \qw & \qw & \qw & \rstick{\ket{a_3}} \gategroup{1}{5}{4}{7}{0em}{--} \gategroup{1}{7}{4}{15}{0em}{--}\\
\lstick{\ket{a_0}=|0\rangle} & \gate{H} &\qw & \ctrl{-3}    &\qw & \ctrl{-1} &  \qw & \qw &\qw &  \qw & \qw & \qw & \qw& \qw & \qw & \ctrl{-1} & \qw  & \ctrl{-3} &\qw &\qw & \qw & \rstick{\ket{a_0}} \gategroup{1}{15}{4}{17}{0em}{--} \gategroup{1}{17}{4}{19}{0em}{--}\\
  \lstick{\textcolor{magenta}{Time \;Slice}} & \mbox{\textcolor{magenta}{\# 1}}  & & \mbox{\textcolor{magenta}{\# 2}}  & & \mbox{\textcolor{magenta}{\# 3}}  & & & & \mbox{\textcolor{magenta}{\# 4}}  & & & & & & \mbox{\textcolor{magenta}{\# 5}} & &  \mbox{\textcolor{magenta}{\# 6}} &  & & & & \\
}
\]
\caption[Gates used to minimize Circuit Family \#$3$ gate depth]{Gates used to minimize Circuit Family \#$3$ gate depth. The precursor and post circuitry used in circuit family \#$3$ for $\mathcal{N}=8$ are for setting up the parallel CSwap execution stages in Figure \ref{fig25}.}
\label{fig26}
\end{center}
\end{figure}

Circuit family \#3 reduces the CSwap gate depth for time slice \#$k$ from $k$ to $Log_2(\,k\,)$. This reduction is implemented by using a tree-like circuit structure which enables {\em parallel} execution of the $k$ CSwap's 
for the $k$'th time slice. The tree-like parallelization circuit takes a circuit of gate depth $Log(k)$ to construct,
thereby limiting the parallelization construction to a gate depth at time slice \#$k$ of $Log_2(k)$.
Figure \ref{fig25} demonstrates the circuit family \#3 approach for the scenario of $\mathcal{N}=8$ classical bits.\footnote{Therefore $n\,=\,Log_2(\mathcal{N})\,=\,Log_2(8) \,= \,3$.} Every gate within a time slice is executed simultaneously. Thus, in time slice \#4 all four controlled swap's are executed in parallel. This is allowed as all four CSwap's act on qubits which are not involved with any other gate {\em in that time slice}. The tradeoff is that a circuit of gate depth $Log_2(k)$ is needed to set up the simultaneous execution of the four CSwap's in one time slice. This $Log$ depth circuitry is shown in time slices \# 2 and \#3 in Figure \ref{fig25}. Focussing on qubits $|a_0,\,a_1,\,a_2,\,a_3\rangle$ and the first three times slices of Figure \ref{fig25} leads to Figure \ref{fig26}, where the CSwap gates shown in time slice \# 4 of Figure \ref{fig25} are removed for clarity. 

\renewcommand{\arraystretch}{1.25}
\begin{table}[h!]
\begin{center}
\begin{tabular}{||c | c ||}
\hline \hline 
 Location in Circuit & Quantum State $|a_0,\,a_1,\,a_2,\,a_3\rangle$ \\
 \hline  \hline
 Before Time Slice \#1  & $|0000\rangle$ \\
 \hline
 After Time Slice \#1 & $(|0\rangle +|1\rangle)\otimes|000\rangle$ \\
  \hline
 After Time Slice \#2 & $(|00\rangle+|11\rangle)\otimes |00\rangle$ \\
  \hline
 After Time Slice \#3 & $|0000\rangle+|1111\rangle$ \\
\hline \hline  
\end{tabular}
\end{center}
\caption[Tracking qubit state transformations in a Circuit Family \#3 layout]{Working through the quantum state transformations among the ancilla qubits in the circuit shown in 
Figure \ref{fig26} for parallelizing the CSwap's execution. The resulting four qubit state $|a_0,a_1,a_2,a_3\rangle $ is a Shor Cat state. Time slices \#5 and \#6 serve to decouple the ancilla qubits 
$a_1$, $a_2$ and $a_3$ from the remaining qubits, allowing $a_1$, $a_2$ and $a_3$ to be used elsewhere in the circuit. {\em Warning: Note that for ease of presentation the sequence of qubits in the ket shown in the rightmost column of Table \ref{table25} is different than the top $\rightarrow$ down sequence of qubits shown in the Figure \ref{fig26} circuit.}}
\label{table25}
\end{table}

\subsection{Gate depth analysis for Circuit Family \#3}

As done earlier, without loss of generality, let $\mathcal{N}=2^n$ be the total number of classical bits to be loaded into 
a quantum state. Circuit family \#3 follows the same general recursion scheme of circuit family \#2, but with
additional ancilla circuitry. As in circuit family \#2, the recursion is broken down into a total of $n$ stages. 
From the circuits in Figures \ref{fig25} and \ref{fig26}, as well as the state transformations
shown in Table \ref{table25}, the gate depth of each of the $k$ stages can be calculated. For stage $k$, where $k \in \{ 1,2,3,\cdots, n\}$, one time slice is dedicated to the Hadamard and one time slice is dedicated to executing {\em all} 
of the stage $k$ CSwap's. In addition, each stage has one ancilla construction circuit and a corresponding ancilla deconstruction circuit. Each of these latter circuits have gate depth $\lceil \; Log_2(k) \;\rceil$. Thus the gate depth of a family \# 3 circuit loading $\mathcal{N}$ classical bits is as given in Equation \ref{family3tally}.

\begin{equation}
Gate \; Depth \,=\, \sum_{k=1}^{k=n} \, \Big ( 1 + 1  + \,\lceil \, Log_2(k) \,\rceil \,\Big ) \;  
\label{family3tally}
\end{equation}

\begin{equation}
\le \; 
\sum_{k=1}^{k=n} \, \Big ( \,3 +\, Log_2(k)  \,\Big ) \; =\; 3\,n \,+\, \sum_{k=1}^{k=n} \,Log_2(k),
\label{family3tallyIII}
\end{equation}

where the bound $\lceil \, Log_2(k) \,\rceil \;\le \;  1 +\, Log_2(k) $ is used. 
The key quantity in Equation \ref{family3tallyIII} is the term $\sum_{k=1}^{k=n} \;Log_2(k)$.
Rewriting the summation term in Equation \ref{family3tallyIII} yields

\begin{equation}
\sum_{k=1}^{k=n}\quad Log_2(k)\quad=\quad Log_2(e) \; Ln\left ( \;\prod_{k=1}^{k=n} \,k\;\right ).
\label{eqn104}
\end{equation}

Since 
$
\;\prod_{k=1}^{k=n} \,k\;=\; n\,!,
$
applying Stirling's approximation to first order yields Equation \ref{eqn105}.\footnote{For Stirling's approximation, see {\it Mathematical Methods in the Physical Sciences} by Mary Boas, Second Edition, Section 11, Page 472, Equation 11.1.} 

\begin{equation}
\sum_{k=1}^{k=n}\quad Log_2(k)\;=\; Log_2(e) \; n \; Ln(\,n\,),
\label{eqn105}
\end{equation}

where $Ln(\,n\,)$ is the natural logarithm of $n$, specifically $Ln(\,n\,)\,\equiv\,Log_e(n)$. 
Rewrite $Ln(n)\,=\,\frac{\;Log_2(n)\;}{\;Log_2(e)\;}$ yielding 

\begin{equation}
\sum_{k=1}^{k=n}\quad Log_2(k)\;=\; n \; Log_2(\,n\,).
\label{eqn106}
\end{equation} 

\begin{figure}[h!]
\begin{center}
\includegraphics[height=7in,width=7in,keepaspectratio=true,angle=0]{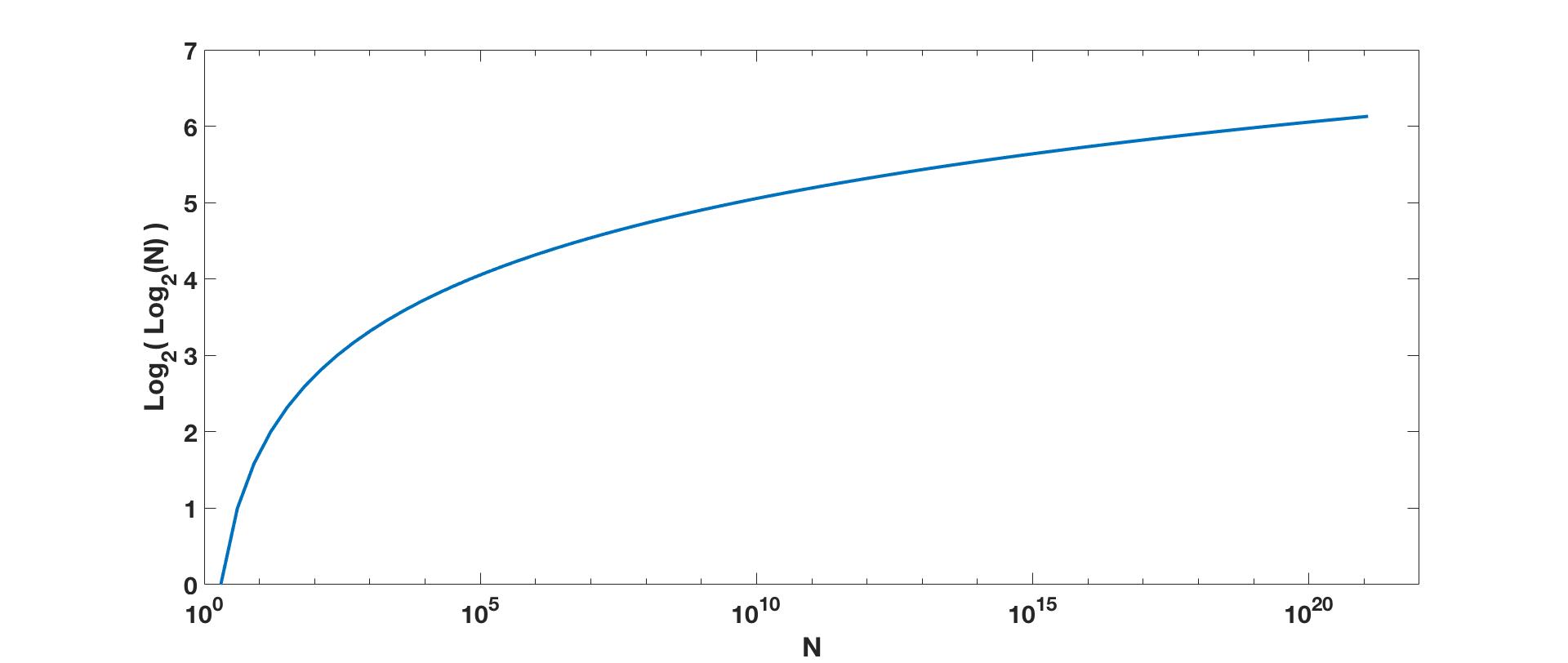}
\caption[$Log_2\Big ( \;Log_2\big(\,\mathcal{N} \,\big)\;\Big)$ versus $\mathcal{N}$ plot]{Plot of $Log_2\Big ( \;Log_2\big(\,\mathcal{N} \,\big)\;\Big)$ versus $\mathcal{N}$. See Equation \ref{eqn107} for application.}
\label{fig19}
\end{center}
\end{figure}

Recall that $n\,=\,Log_2(\mathcal{N})$ and one obtains 

\begin{equation}
\sum_{k=1}^{k=n}\quad Log_2(k)\;=\; Log_2(\,\mathcal{N}\,) \;\;\; Log_2\Big (\,Log_2(\,\mathcal{N}\,)\,\Big )\;\approx\;Log_2(\,\mathcal{N}\,)
\label{eqn107}
\end{equation} 

where the justification for the approximation in Equation \ref{eqn107} is given by the plot of $Log_2\Big (\,Log_2(\,\mathcal{N}\,)\,\Big )$ shown in Figure \ref{fig19}.
Using the result from Equation \ref{eqn107} in Equation \ref{family3tallyII} yields the gate depth of 
circuit family \#3.


\begin{equation}
Gate \; Depth \; =\; 3\,n \,+\, \sum_{k=1}^{k=n} \,Log_2(k)\;\approx\; \;\mathcal{O}\Big (\,Log\left (\,\mathcal{N}\,\right)\,\Big ),
\label{family3tallyII}
\end{equation}

where the fact that $Log_2\Big (\,Log_2(\,\mathcal{N}\,)\,\Big )$ grows very slowly with 
increasing $\mathcal{N}$, as shown in the plot of Figure \ref{fig19}, is used. Note that $Log_2\Big (\,Log_2(\,\mathcal{N}\,)\,\Big )$ is essentially 
a constant less than $10$ in value for large, but finite, $\mathcal{N}$ in 
the numerical range of interest for the data loading circuitry.  

\newpage

\newpage 

\section{Summary}

All the circuit families discussed in this document, in both the erasure and no erasure forms, have been simulated in the Quipper quantum computer simulation framework. Quipper is a well known classical computer based 
software tool used by the quantum  computing community
as a test and verification framework for proposed quantum circuits and algorithms. 

To summarize, the work presented in this document describes the following.

\begin{itemize}
\item
A practical circuit family (\#3) which loads $\mathcal{N}$ classical bits into a quantum data structure of size $Log_2(\mathcal{N})$ qubits in 
a quantum circuit depth of $\mathcal{O}\Big (Log(\mathcal{N})\Big )$. Both this data structure size and this gate depth 
are critical requirements for generic quantum algorithms and circuits exhibiting exponential speedup over their classical algorithm and circuit counterparts. 
\item
Classical compression with quantum decompression can ease the 
complexity and gate count of the data loading circuitry, while optimizing the transfer of bits into qubits. 
The design methodology described shows how classical compression/quantum 
decompression schemes can be designed 
using classical compression and decompression algorithms. 
\item
Ancilla qubits are used in circuits detailed in this document, as well as in most circuits in the 
literature. Ancilla qubits are typically discarded after use, which means that in practice the 
ancilla qubits are preserved, but ignored, until the end of the computation\cite{watrous1}. In 
practice, rather than preserve but ignore these idle qubits until the completion of the computation, 
one would like to reuse these ancilla qubits so as to keep the spatial overhead (the \# of qubits 
used by the computation and/or circuit) to a minimum.  
Section \ref{reusequbits} of this document describes a procedure and associated family of 
quantum circuits 
which decouple discarded qubits from the main body of qubits in use during the computation, 
thereby enabling the discarded qubits to be reused later in the circuit and as a result 
minimizing overall spatial 
qubit usage in a quantum circuit computation. 
\end{itemize}

\section{Acknowledgements}

The authors would like to thank Professor Aram Harrow of the Massachusetts Institute of Technology Center for Theoretical Physics for stimulating discussions.


\section{Bibliography}
\begin{thebibliography}{5}


\bibitem{watrous1}
{\em An Introduction to quantum information and quantum circuits} by John Watrous, Published in The Association of Computing Machinery - Special Interest Group in Algorithms and Computation Theory (ACM SIGACT), June, 2011, Volume 42, Issue Number 2, Page 52-67.


\end{thebibliography}
\end{document}